\documentclass[sn-chicago]{sn-jnl}

\jyear{2023}%

\theoremstyle{thmstyleone}%
\theoremstyle{thmstyletwo}%

\theoremstyle{thmstylethree}%

\usepackage{amsmath}

\usepackage{color}

\newcommand{\FinalModif}[1]{{ {{{#1}}}}}

\newcommand{\aj}{AJ}
\newcommand{\apj}{ApJ}
\newcommand{\apjs}{ApJ Suppl. Series}
\newcommand{\apjl}{ApJL}
\newcommand{\mnras}{MNRAS}
\newcommand{\aap}{A{\&}A}
\newcommand{\nat}{Nature}
\newcommand{\actaa}{Acta Astronomica}
\newcommand{\solphys}{Solar Physics}
\newcommand{\ssr}{Space Science Reviews}
\newcommand{\pre}{Physical Review E}
\newcommand{\araa}{Annual Review of Astron and Astrophys}

\usepackage{hyperref}
\hypersetup{
    colorlinks=true,
    linkcolor=blue,
    filecolor=magenta,      
    urlcolor=cyan,
    pdftitle={Overleaf Example},
    pdfpagemode=FullScreen,
    }

\newcommand{\Pra}{{\rm Pr}}
\newcommand{\Pm}{{\rm Pm}}

\begin{document}

\title[Dynamics of the tachocline]{Dynamics of the tachocline}

\author[1]{\fnm{Antoine} \sur{Strugarek}}\email{antoine.strugarek@cea.fr}
\author*[2,5,7]{\fnm{Bernadett} \sur{Belucz}}\email{b.belucz@sheffield.ac.uk}
\author[1]{\fnm{Allan Sacha} \sur{Brun}}\email{sacha.brun@cea.fr}
\author[3]{\fnm{Mausumi} \sur{Dikpati}}\email{dikpati@ucar.edu}
\author[4,6]{\fnm{Gustavo} \sur{Guerrero}}\email{guerrero@fisica.ufmg.br}

\affil[1]{Université Paris-Saclay, Université Paris Cité, CEA, CNRS, AIM, 91191, Gif-sur-Yvette, France}

\affil[2]{Solar Physics and Space Plasma Research Center, School of Mathematics and Statistics, University of Sheffield, S3 7RH, UK}

\affil[3]{High Altitude Observatory, NCAR, 3080 Center Green Drive, Boulder,
80301, CO, USA}

\affil[4]{Universidade Federal de Minas Gerais,  Av. Pres. Antônio Carlos, 6627, Belo Horizonte - MG, 31270-901}

\affil[5]{Hungarian Solar Physics Foundation, Gyula, Hungary}

\affil[6]{New Jersey Institute of Technology, Newark, NJ 07103, USA}

\affil[7]{Department of Astronomy,Institute of Geography and Earth Sciences, E\"otv\"os University, Budapest, Hungary}

\abstract{The solar tachocline is an internal region of the Sun possessing strong radial and latitudinal shears straddling the base of the convective envelope. Based on helioseismic inversions, the tachocline is known to be thin (less than 5\% of the solar radius). Since the first theory of the solar tachocline in 1992, this thinness has not ceased to puzzle solar physicists. In this review, we lay out the grounds of our understanding of this fascinating region of the solar interior. We detail the various physical mechanisms at stake in the solar tachocline, and put a particular focus on the mechanisms that have been proposed to explain its thinness. We also examine the full range of MHD processes including waves and instabilies that are likely to occur in the tachocline, as well as their possible connection with active region patterns observed at the surface. We reflect on the most recent findings for each of them, and highlight the physical understanding that is still missing and that would allow the research community to understand, in a generic sense, how the solar tachocline and stellar tachocline are formed, are sustained, and evolve on secular timescales.}

\keywords{solar tachocline, magnetohydrodynamics of stars, shear and magnetic instabilities, solar dynamo}



\maketitle

\section{The solar tachocline}
\label{sec:solar-tach-import}

\subsection{What is known about the solar tachocline?}
\label{sec:observ-solar-tach}

\begin{figure}
    \centering
    \includegraphics[width=\linewidth]{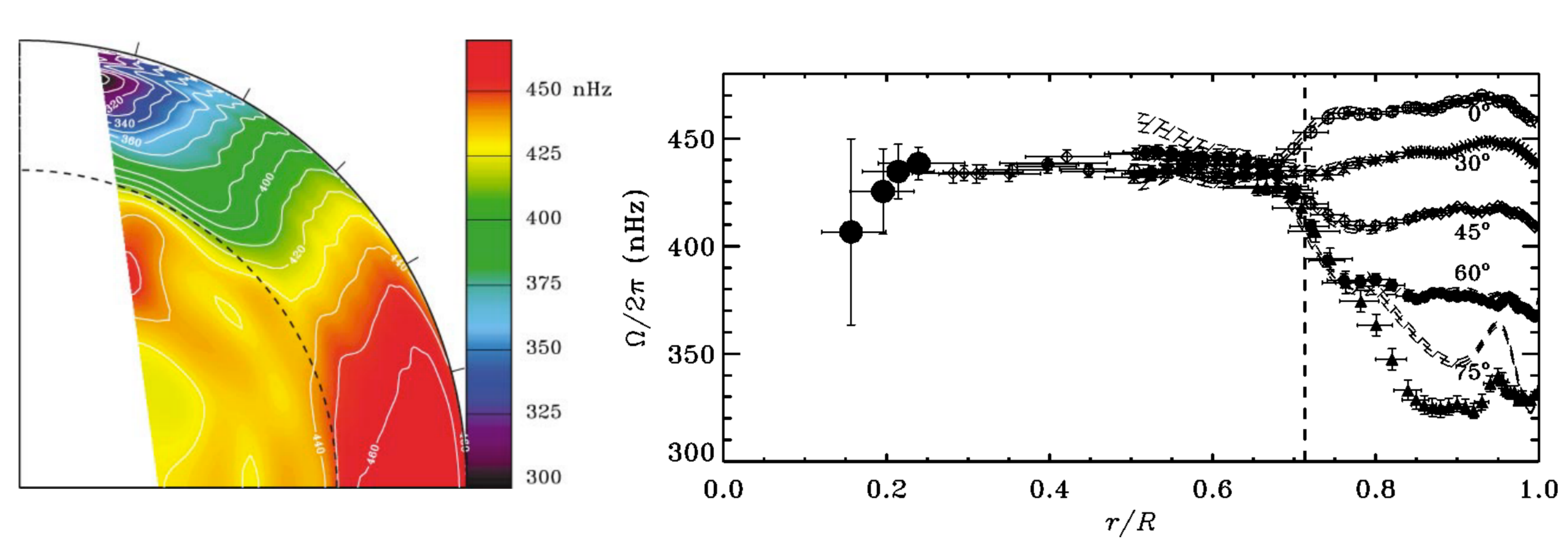}
    \caption{\textit{Left}: Solar differential rotation, as deduced from
      helioseismology \citep{1998ApJ...505..390S,2003ARA&A..41..599T}. The surface exhibits a
      shear layer (often dubbed \textit{near-surface shear
        layer}). The bulk of the convective envelope shows a
      latitudinal differential rotation with mostly conical
      iso-contours, except perhaps at the equator where contours are
      more aligned with the rotation axis. At the base of the
    convection zone, the rotation becomes progressively solid, creating
  a sheared layer called the tachocline. \textit{Right}: Radial cuts
  of the solar differential rotation profile at latitudes 0$^\circ$,
  30$^\circ$, 45$^\circ$, 60$^\circ$ and 75$^\circ$ (the dashed line
  shows the results with a different inversion technique using
  regularized least-squares, see \citet{2003ARA&A..41..599T} and references
therein for more details).}
    \label{fig:DRThompson}
\end{figure}

After several decades of observations, one of the great achievements
of helioseismology remains the inversion of the internal rotational
profile of the Sun \citep{1989ApJ...343..526B}. The surface differential
rotation prevails through the whole convective zone, as seen in Figure \ref{fig:DRThompson}
\citep{2003ARA&A..41..599T}. In the radiative interior, the rotation rate is
uniform and matches the rotation rate of the convective envelope at
mid-latitudes (about 430 nHz near latitude $\pm$35$^\circ$). The transition
from a differential to uniform rotation occurs in a transition layer
known as the tachocline \citep{2007sota.book.....H}. \FinalModif{This region lies just beneath the convection zone and likely has a prolate form, i.e., it is located at $\sim0.693 R_{\odot}$ close the equator, and at $\sim 0.717 R_{\odot}$ at higher latitudes \citep{1999ApJ...527..445C, 2003ApJ...585..553B}.  Its thickness slightly varies according to the definition considered in the helioseismic forward modeling determinations \cite[see, for instance][]{1996ApJ...469L..61K,1999A&A...344..696C,1999ApJ...516..475E, 2001MNRAS.324..498B}, yet it is certainly very thin (less than 5\% of the solar radius). Recent results considering $\sim 20$ years of observations indicate that neither the position or the thickness of the tachocline significantly change with the solar cycle \citep{2019ApJ...883...93B}.    The tachocline is subject to strong latitudinal and radial shears. The radial shear in the mean azimuthal flow is the strongest (by almost an order of magnitude) and changes sign at latitude $\pm$35$^\circ$. Both the radial and the latitudinal shears exhibit significant changes in their amplitudes with the solar cycle. Interestingly, these changes have been found to be different between the cycles 23 and 24 \citep{2019ApJ...883...93B}. }  

This strong  large-scale shear drew the attention to the tachocline, and led it to be considered a prominent
player in many dynamo models of the Sun known as 'interface dynamos' \citep{1993ApJ...408..707P}. Indeed, such a strong large-scale shear  is a very efficient converter of poloidal fields into toroidal field in mean-field models (see Chapters 1, 12 and 15). Nevertheless,
the real role played by the tachocline in solar and stellar dynamos is still controversial.  
On one hand, ZDI observations suggest that the magnetic topology changes from a simple dipolar structure for young TTauri, fully-convective, objects to a complex multipolar topology for objects that have already developed a radiative core \citep{2012ApJ...755...97G}. 
On the other hand, \FinalModif{following earlier works from \citet{2003ApJ...583..451M,2012AJ....143...93R},}
\citet{2016Natur.535..526W} \FinalModif{unambiguously} showed that the X-ray luminosity of fully-convective stars
followed the same empirical trend than slightly hotter stars with a
radiative core, showing no strong difference in the case a tachocline
could be present. In addition, \citet{2016ApJ...830L..27R} discovered magnetic
activity cycles in ultra-cool dwarfs which are fully-convective, again
showing seemingly no strong impact of the existence or non-existence of a
tachocline. In the theoretical field, \FinalModif{global numerical simulations (e.g. \citealt{2017Sci...357..185S}) also showed}
that a tachocline was actually not necessary to
obtain Sun-like cyclic dynamos, and \citet{2020ApJ...893..107B} 
showed
that a tachocline was not necessary to generate strong wreaths of toroidal field 
within the convective core of M-dwarfs. It could nevertheless influence how
the dynamo operates. Indeed, numerical experiments show that
instabilities in the tachocline could affect the magnetic cycle
period \citep{2015ApJ...813...95L}, by modulating (due to instabilities) the Poynting flux
permeating the base of the convective envelope (more on this in Section \ref{sec:inst-solar-tach}). More recently
\citet{2019ApJ...880....6G} also reported a similar instability in their
numerical endeavors and \citet{2022ApJ...926...21B} found hints in their simulations that taking the tachocline into account could affect the cycle period \citep[see also][]{2018ApJ...859...61B}.

\subsection{Why is it so thin?}
\label{sec:why-it-so}

This simple question about the solar tachocline has puzzled and fascinated solar physicists since its
discovery. Indeed, its extreme thinness remains one of the key
mysteries of our star. Right after its discovery,
\cite{1992A&A...265..106S} produced the first hydrodynamic theory of
the solar tachocline. They analyzed the long-term equilibrium of this
thin shear layer, and showed that it should burrow on secular
time-scales due to the phenomenon of \textit{radiative spreading}. In
a nutshell, the latitudinal shear 
with conical iso-contours
requires an associated temperature
gradient to break the Taylor-Proudman constraint. A meridional
circulation then follows, spreading the latitudinal shear downwards
on an Eddington-Sweet timescale, defined as
  \begin{equation}
    \label{eq:tES}
    t_{\rm ES} = \frac{N^2}{\Omega_\odot^2} \frac{R^2}{\kappa}\, ,
  \end{equation}
where the square of the Brunt-Väisälä
frequency is defined as $N^2= g \left(\frac{1}{\gamma} \frac{{\rm d} \ln
    P}{{\rm d} r} - \frac{{\rm d} \ln
    \rho}{{\rm d} r}  \right)$, $\Omega_\odot$ is the mean rotation of
the Sun, $R$ the radius of the tachocline and $\kappa$ the thermal
diffusion coefficient. The Eddington-Sweet timescale based on molecular thermal diffusion reaches about $10^{20}$ years for the solar tachocline \citep{2006A&A...457..665B}.
  
The radiative spreading (or burrowing) has been revisited multiple
times
ever
since
\citep{2008ApJ...674..498G,2009ApJ...704....1G,2012ApJ...755...99W,2018ApJ...853...97W},
confirming essentially the robustness of the mechanism highlighted by
\citet{1992A&A...265..106S}. A back-of-the-envelope calculation quickly shows
that according to this theory, the solar tachocline should have spread
by at least $0.3 R_\odot$ into the radiative zone at the solar
age. Hence, additional mechanisms must be at play to confine the solar
tachocline to its extreme thinness smaller than 5\% $R_\odot$.

Multiple physical scenarios have been proposed to confine the solar
tachocline. They can be decomposed into two families: the fast
tachocline scenarios, relying on physical mechanisms on timescales of
months to years; and the slow tachocline scenarios, relying on physical
mechanisms acting on timescales longer than a millennium. The large
community debate about the solar tachocline confinement can therefore be largely attributed to
the difficulty to adequately model it, because of the multiple
time-scales involved in its dynamics and evolution. 

\begin{figure}
    \centering
    \includegraphics[width=\linewidth]{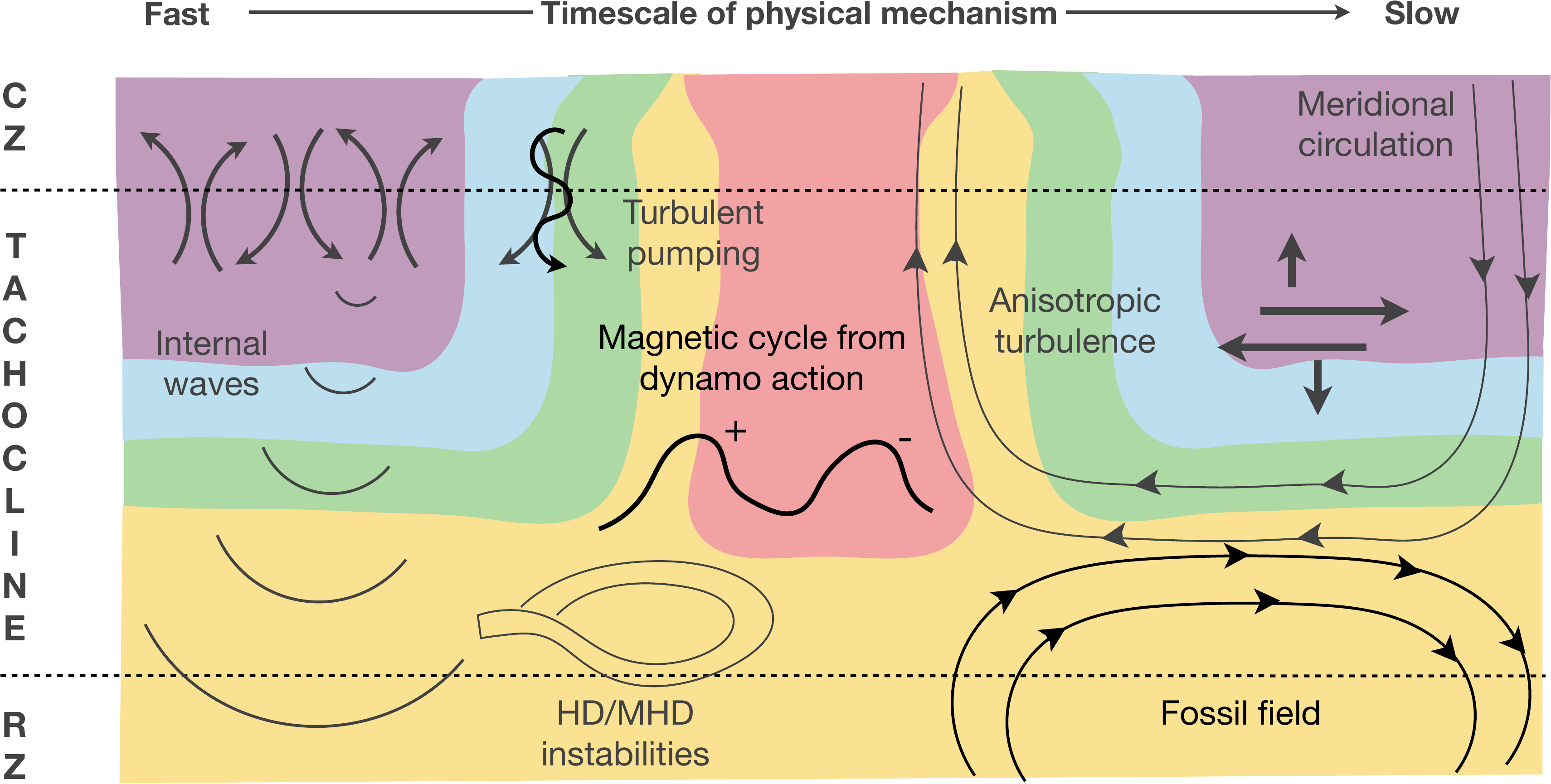}
    \caption{Schematic of the physical processes at play in the various proposed scenarios to confine the tachocline. The typical timescale of the physical process increases from left to right. The solar differential rotation in the tachocline is illustrated by the background colored zones (see also Fig. \ref{fig:DRThompson}).}
    \label{fig:schematic}
\end{figure}

The multiple physical mechanism behind confinement scenarios proposed to explain the thinness of the solar
tachocline are illustrated in Figure \ref{fig:schematic} and ordered as a function of their physical timescale (the differentiation between slow and first processes in the tachocline was first described by \citealt{2000SoPh..192...27G}, see e.g. the Table 1 there). The various processes can be summarized as follows (we will dive deeper into the
last three in Section \ref{sec:magn-conf-scen} and Section \ref{sec:inst-solar-tach}). 
\begin{itemize}
\item \textbf{\textit{Anisotropic viscosity and turbulent transport}}. It was originally proposed by
  \citet{1992A&A...265..106S} that turbulent viscosity in the upper tachocline
  was likely to be anisotropic due to the strong stratification
  encountered by
  convective plumes penetrating from above. They modelled it by simply
  considering a dominant horizontal viscosity, and in this case the tachocline could be
  confined to its present thickness, provided the effective (turbulent) Prandtl
  number was large enough. This scenario was opposed by
  \citet{1998Natur.394..755G}, who argued that turbulence was unlikely to
  transport angular momentum in a diffusive manner on such a quasi-2D
  layer, basing their argument on what occurs in the Earth atmosphere. In fact, this debate is not settled today. In a series of
  analytical
  papers \citep{2005A&A...441..763K,2006A&A...456..617L,2007A&A...468.1025K,2009PhRvE..80b6302L}, Kim and Leprovost showed quite clearly that depending on
  the considered characteristics of the solar tachocline (horizontal shear,
  radial shear, stratification, magnetism, waves) turbulent transport could
  actually be either diffusive or anti-diffusive. These different behaviours were confirmed to exist with numerical simulations by \citet{2003ApJ...586..663M}. Indeed, the consideration of low-{\Pra} physics changes the situation when compared to the Earth
  atmosphere, making the direction and amplitude of the turbulent transport in the tachocline not so
  obvious to estimate \citep{2020ApJ...901..146G}. Furthermore,
  \citet{2007ApJ...667L.113T} showed that in the presence of large-scale
  magnetic fields, turbulence could sometimes even not transport angular
  momentum efficiently at all. This debate was further pushed forward
   recently by \citet{2020ApJ...892...24C}, who developed a more complete
  view of the problem and proposed that the tachocline could be
  confined by the combined effect of the relaxation of potential
  vorticity gradient along a resisto-elastic drag originating from
  tangled magnetic fields in the solar tachocline. Unsurprisingly, turbulent transport
  therefore still remains one of the most promising and debated origin
  for the solar tachocline confinement.
\item \textbf{\textit{Magnetic confinement with a fossil
      field}}.
      Following \citet{1992A&A...265..106S}, \citet{1998Natur.394..755G} wrote
  a very influential paper on the confinement of the solar
  tachocline, after a seminal idea by \citet{1997AN....318..273R} who proposed a confinement of the tachocline due to the existence of a fossil magnetic field. Arguing that turbulent transport could not behave
  like a diffusion (see the previous point for a more critical view on
  this argument), they looked for another possible source for the
  tachocline confinement. They naturally turned to magnetism. In their
  seminal paper, they proposed that a magnetic field of fossil origin
  confined inside the solar radiation zone could balance out the
  inward radiative spreading of the tachocline through Lorentz
  forces. Nevertheless, magnetic fields diffuse on Ohmic, secular
  timescales and an additional mechanism was also required to confine
  the magnetic field itself, leading to a \textit{double-confinement}
  scenario. The large-scale meridional flow accompanying \textit{a
    priori} the burrowing of the tachocline \citep{2012ApJ...755...99W} was in
  turn invoked to play this confining role, leading to an overall
  balance of the solar tachocline. The \citet{1998Natur.394..755G} scenario
  possesses an appealing advantage: it offers a truly large-scale theory
  for the confinement of the solar tachocline, that does not depend on
  the detailed behavior of complex, turbulent MHD flows at the top of
  the solar tachocline. Nevertheless, one potential problem with this scenario 
  lies in the possibility for the Sun to possess of a magnetic-field zone between the top of the convection and the confinement layer where the fossil field 
  Lorentz force is at play. This complex scenario was then questioned with
  various series of numerical simulations (\textit{e.g.} \citealt{2011A&A...532A..34S}, \citealt{2013MNRAS.434..720A}). 
  We will come back to this active debate in Section   \ref{sec:conf-with-foss},
  along with the most recent numerical endeavors on this topic. 
\item \textbf{\textit{Magnetic confinement with a dynamo
      field}}. The dynamo magnetic field of the Sun visible at its
  photosphere certainly permeates deep inside the solar interior. At
  the very least, down to the solar tachocline along with the
  differential rotation and the turbulent convective motions. Many
  dynamo theories have historically given the tachocline an important
  role (see e.g. Chapter 15). If the cyclic magnetic field
  of the Sun permeates the upper tachocline, could it contribute to
  confining it to its observed thinness? This interesting question was
  first addressed by \citet{2001SoPh..203..195F} and further developed in
  a series of papers by the same authors. This scenario historically
  deeply interested Jean-Paul Zahn, who developed an extension of the
  initial development by \citet{2001SoPh..203..195F}. This work was later refined and published in \citet{2017A&A...601A..47B} (more on the details in Section \ref{sec:conf-with-dynamo}).
\item \textbf{\textit{Magneto-hydrodynamic instabilities}}. As a
  sheared, stratified, magnetized thin layer, one may naturally
  question the global stability of the solar tachocline, and whether
  any MHD instabilities could actually help to explain its thinness (\textit{e.g.} \citealt{1981GApFD..16..285W, 1997ApJ...484..439G, 1999ApJ...526..523C}). 
  A general introduction to this topic can be found in section 8.2 of \citet{2005LRSP....2....1M}. 
\end{itemize}

\FinalModif{We now first turn to the magnetic tachocline confinement problem (Section \ref{sec:magn-conf-scen}) and move on to the detailed description of the magneto-hydrodynamic instabilities in the tachocline and their impact on our understanding of the tachocline of the solar magnetism in Section \ref{sec:inst-solar-tach}.}

\section{Magnetic confinement scenarios of the tachocline}
\label{sec:magn-conf-scen}

We now give a detailed description of two mechanisms that have been proposed to confine the tachocline to its present thickness, based on the existence of internal magnetic fields either generated by dynamo processes (Section \ref{sec:conf-with-dynamo}) or of fossil origin (Section \ref{sec:conf-with-foss}).


\subsection{Confinement with a dynamo field}
\label{sec:conf-with-dynamo}

The solar magnetic cycle of 11 years, visible at the photosphere, sustains strong magnetic fields inside
our star. But how strong? It is known for a fact that at the surface of
the Sun, the dipolar magnetic field has a typical value of a few Gauss over the
solar surface \citep{2012ApJ...757...96D}. Moreover, magnetic flux
concentrations in the form of sunspots can reach
magnetic field intensities of \FinalModif{hundreds of kG}. Deeper down in the
solar convective envelope, where the overall dynamo is supposed to be
seated, we do not have any precise observational constraints on the
field amplitude. \citet{2000A&A...360..335A} have tried to derive an upper limit
for the magnetic field amplitude through helioseismic inversion and
found that it should be below 30 T -- a somewhat large upper
value. Models of flux emergence from the base of the convection zone
leading to realistic flux concentration at the photosphere yield a
toroidal field of the order of $1-10$ T
\citep{2009LRSP....6....4F,2018ApJ...857...83J}.

The possibility of confining the solar tachocline with
the oscillating dynamo field penetrating from the upper convective
envelope was first suggested by \citet{2001SoPh..203..195F}, and further explored by \citet{2017A&A...601A..47B}. A minimal 1D mean-field model for this scenario,
which can be mathematically written as

\begin{align}
  \label{eq:EqA_Tacho}
  \partial_\tau a &= \partial^2_{\xi\xi} a \, ,\\ 
  \label{eq:EqB_Tacho}
  \partial_\tau b &= \partial^2_{\xi\xi} b - \left(k\delta\right)^2 b -
                    C_A a u \, , \\
  \label{eq:EqU_Tacho}
  \partial_\tau u &= \Pm \left (\partial^2_{\xi\xi} u -
                    \left(k\delta\right)^2 u \right) +
                    C_L a b \, ,
\end{align}
where $\xi$ is the depth below the bottom of the convection zone.
This model traces the evolution of the azimuthally-averaged poloidal magnetic
field $a$, toroidal magnetic field $b$, and differential rotation
$u$, over a domain of depth $\delta$ below the convection zone. For the sake of simplicity, $a$ is assumed to be constant in
latitude, while the latitudinal shape of the toroidal field and of the
differential rotation is characterized by a unique wave number $k$. The
fields are coupled through the induction term $C_A a u$ in Eq. \ref{eq:EqB_Tacho}, and the
Lorentz force $C_L a b$ in Eq. \ref{eq:EqU_Tacho}. This simplified model allows assessing 
whether the Lorentz force can confine the tachocline by preventing the
downward spread of the differential rotation into the underlying
radiative interior. In this work, $C_A$, $C_L$ and $\Pm=\nu/\eta$
(the magnetic Prandtl number) were varied to assess in which parameter regime a
confinement was realized.

\begin{figure}
  \begin{minipage}{0.49\textwidth}
    \centering
    \includegraphics[width=\linewidth]{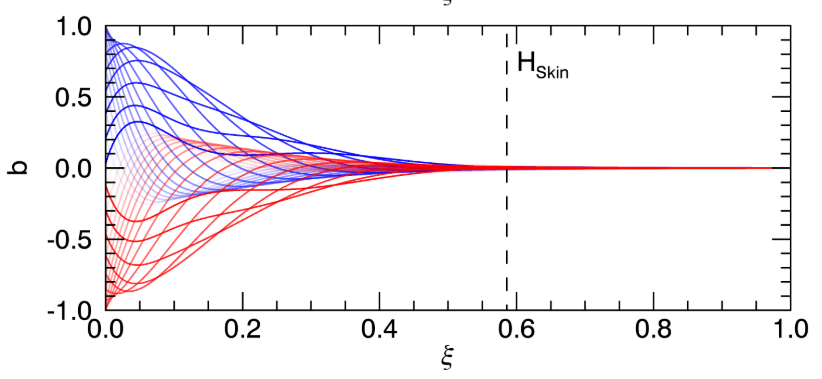}
    \includegraphics[width=\linewidth]{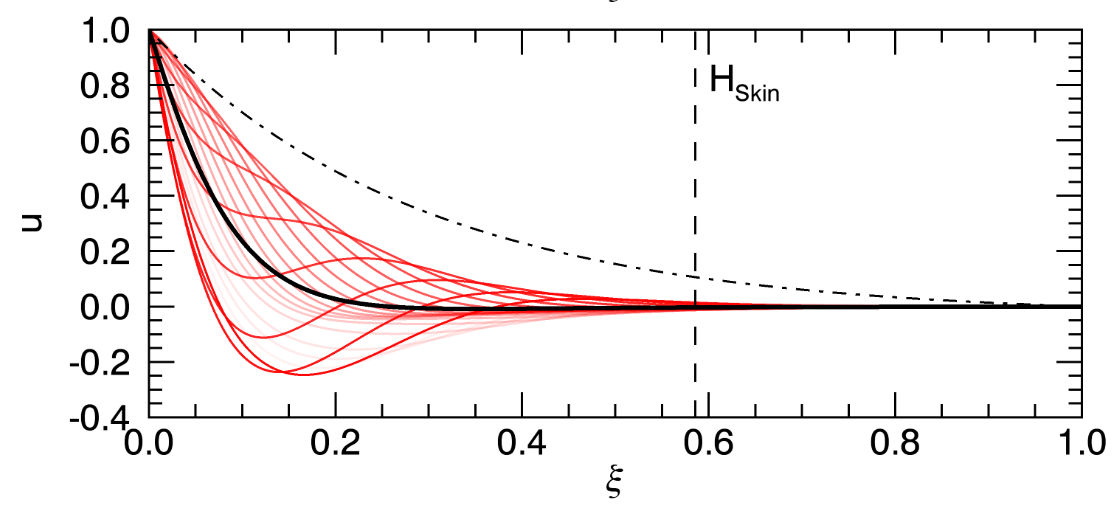}
  \end{minipage}\hfill
  \begin{minipage}{0.49\textwidth}
    \centering
    \includegraphics[width=\linewidth]{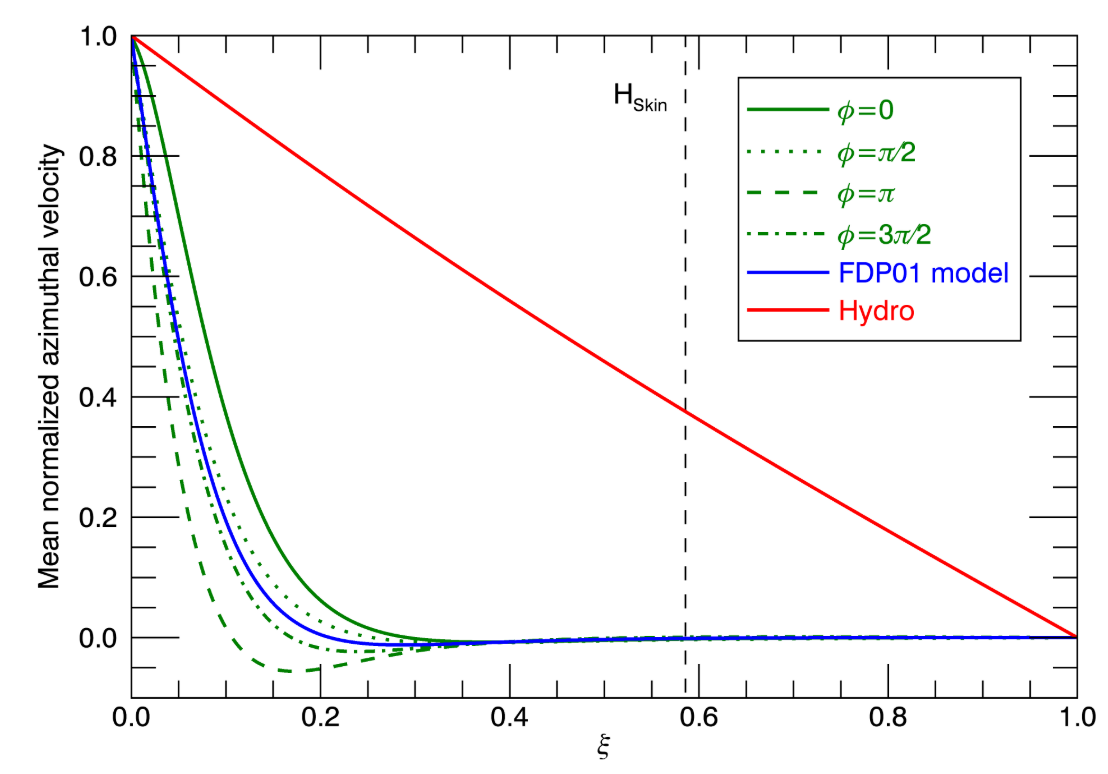}
    \end{minipage}
    \caption{\textit{Left panels}: Original confinement scenario of
      \citet{2001SoPh..203..195F} as revisted by \citet{2017A&A...601A..47B}. The upper panel shows the toroidal
      field $b$, and the lower panel the differential rotation u, as a
    function of depth $\xi$ from left to right. The skin depth $H_{\rm
      skin}=\sqrt{2\eta P_{\rm cyc}}$ is shown by the vertical dashed line. The color and transparency
    of the line label different phases in one magnetic
    cycle. \textit{Right panel}: Study of the influence of the phasing
  between the two components $a$ and $b$ of the magnetic field. The
  angle $\phi$ corresponds to this dephasing. The blue curve is the
  reference model of \citet{2001SoPh..203..195F} that differs from the
  green models in the boundary condition imposed on $a$. A reference
  hydrodynamic model is shown by the red curve. All
  models considering a dynamo cycle (red and green curve) manage to
  confine the tachocline regardless of the assumed relative phase $\phi$.}
    \label{fig:BarnabeSol}
\end{figure}

Let's consider first the original scenario proposed by
\citet{2001SoPh..203..195F} where an oscillating magnetic field \FinalModif{is imposed} at
the top of the domain with a period $P_{\rm cyc}=22$ years, with
the poloidal and toroidal fields being out of phase by $\pi/2$.
The two left panels of
Figure \ref{fig:BarnabeSol} show the evolution of the toroidal field (upper panel)
and of the differential rotation (lower panel). As expected, the
toroidal field oscillates due to the boundary forcing and penetrates
down to the electro-magnetic skin-depth $H_{\rm skin} =
\sqrt{2\eta P_{\rm cyc}}$. The differential rotation $u$ penetrates
over about the same depth, as seen in the lower panel (the average over
$P_{\rm cyc}$ is shown by the thick black line). In this case, the tachocline
is indeed confined, as the initial profile shown by the dash-dotted
line was actually penetrating deeper down than the converged rotation profile. 

The first criticism of the work of \citet{2001SoPh..203..195F} was the
assumed phase between the toroidal and poloidal magnetic
field. Indeed, this phasing can play a critical role to ensure that
the Lorentz force is actually strong enough to confine the
tachocline. This hypothesis was alleviated in \citet{2017A&A...601A..47B} (see Figure \ref{fig:BarnabeSol}), who found that the phasing of the two components of the magnetic field actually marginally
affect the net differential rotation. In all cases (shown in green),
the tachocline remains confined, with a somewhat different radial
profile. The conclusion of this work is that this confinement scenario is actually
quite robust with respect to the relative phase between the poloidal and toroidal components of the dynamo field.

The second and more important criticism 
to this
scenario lies in its
simplicity. Indeed, the original scenario did not consider that the transport in
tachocline could be either diffusive or anti-diffusive, or that it could be
 subject to a radiative spreading (see
Section \ref{sec:why-it-so}). Again, this limitation was also alleviated in the work of \citet{2017A&A...601A..47B} who identified the region in parameter space where the confinement could be strong enough to oppose such spreading mechanisms. Interestingly, these results show that the dynamo-based confinement scenario seems robust for a large variety of turbulent transport scenarios.    

\begin{figure}
  \centering
    \includegraphics[width=0.6\linewidth]{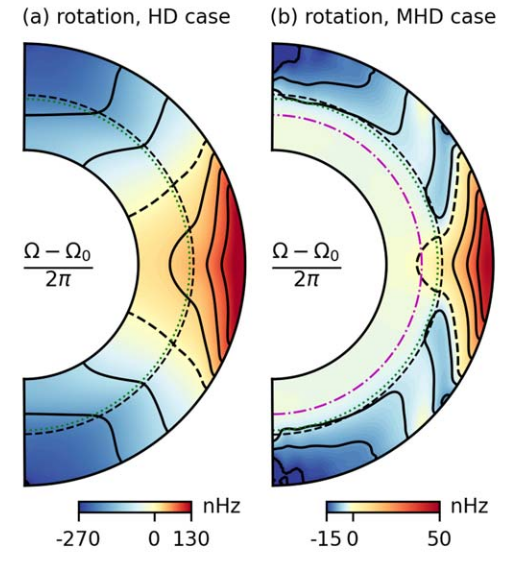}
    \caption{Magnetic field confinement due to a convective dynamo generated in the global simulations of \cite{2022ApJ...940L..50M}. Panel (a) shows viscous tachocline spreading in a hydrodynamic simulation. Panel (b) shows the formation of a tachocline and a solid body rotating radiative interior in a MHD simulation.}
    \label{fig:Matilsky22a}
\end{figure}

The confinment due to a dynamo generated magnetic field has furthermore been recently confirmed by global simulations \citep{2022ApJ...940L..50M}.  Their model considers a fraction of a stable layer and a convection zone with a thermodynamic stratification that resembles the solar interior up to $0.947 R_{\odot}$ and rotates 3 times faster than the Sun which allows to get fast equator and slower poles.  If the simulation does not include magnetic field, the angular momentum spreads from the convection zone to the stable interior because of the viscosity. If the simulation includes magnetic field, a dynamo develops generating non-axisymmetric wreaths of toroidal field around the equator (compare left and right panels of Figure~\ref{fig:Matilsky22a}). This field has a small phase difference with the poloidal magnetic field. This phase difference is instrumental in generating axisymmetric large-scale Maxwell stresses which balances the viscous stresses preventing the inwards transport of angular momentum.  There is a skin-depth of penetration of the magnetic field in the stable layers such as in the models of \cite{2001SoPh..203..195F,2017A&A...601A..47B}. 

In such a confinement scenario, the differential rotation exhibits a
variability on the magnetic cycle timescale, over the electro-magnetic
skin-depth. Such strong variability was not systematically detected in
the Sun \citep{2011JPhCS.271a2075H}. The spatial resolution limit of helioseismic
inversions is about $5\%$ of the solar radius \citep{2009LRSP....6....1H}. If the
tachocline is confined by the oscillating dynamo field, this yields an
upper limit of about $5 \times 10^{10}$ cm$^2$/s for the Ohmic
diffusivity \citep{2017A&A...601A..47B}. 

One can summarize the present understanding of the tachocline confinement scenario by an oscillating dynamo magnetic field as follow:
\begin{itemize}
\item The confinement scenario based on the cyclic dynamo field of the
  Sun is a robust scenario with respect to the various hypothetical
  physical ingredients participating in the angular momentum transport
\item For realistic solar parameters, this confinement scenario works
  only if the magnetic field is able to permeate sufficiently
  deep. In other words, if the turbulent Ohmic diffusivity in the
  upper tachocline exceeds $10^{8}$ cm$^2$/s, it is very likely that 
  the cyclic, large scale field of the Sun is able to confine the
  tachocline to a thickness smaller than $5\%$ of the solar radius
\item 
The plausibility of this confinement scenario has been confirmed through global MHD simulations of convective dynamos \citep{2022ApJ...940L..50M}.  Although the parameter regime of these simulations is still far from realistic, it is an encouraging alternative to explore in future  simulations.
\end{itemize}

\subsection{Confinement with a fossil field}
\label{sec:conf-with-foss}

\begin{figure}
  \centering
  \includegraphics[width=\linewidth]{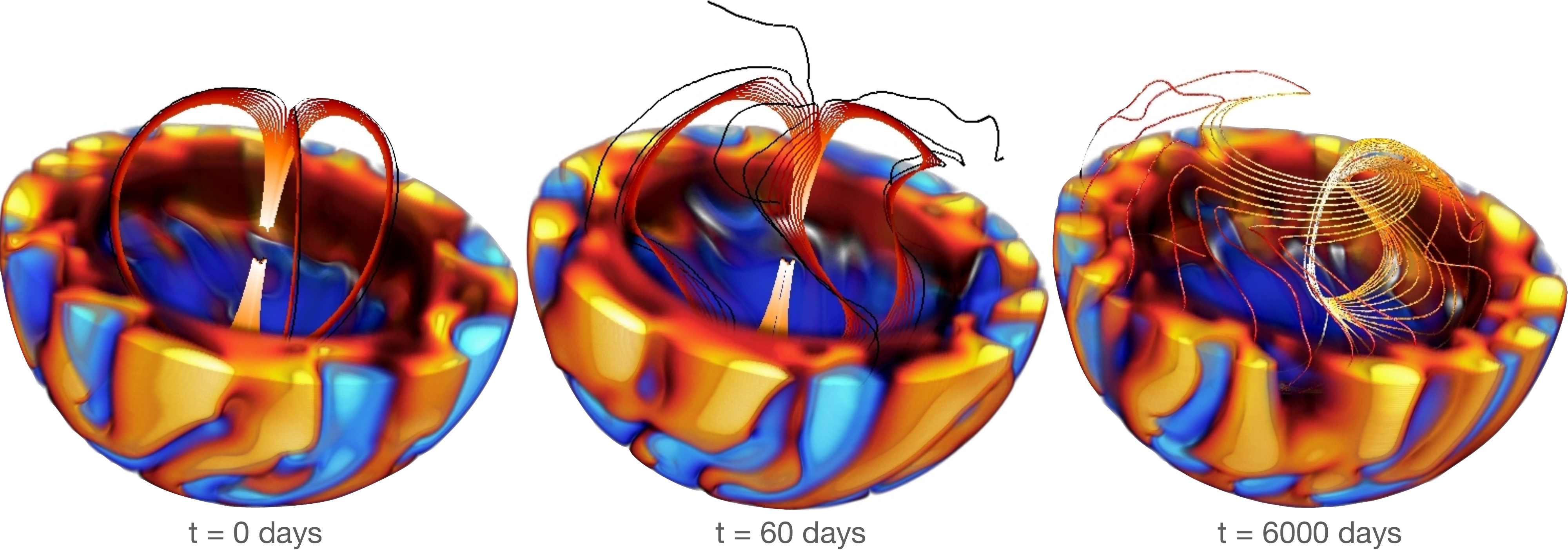} \\
  \includegraphics[width=0.5\linewidth]{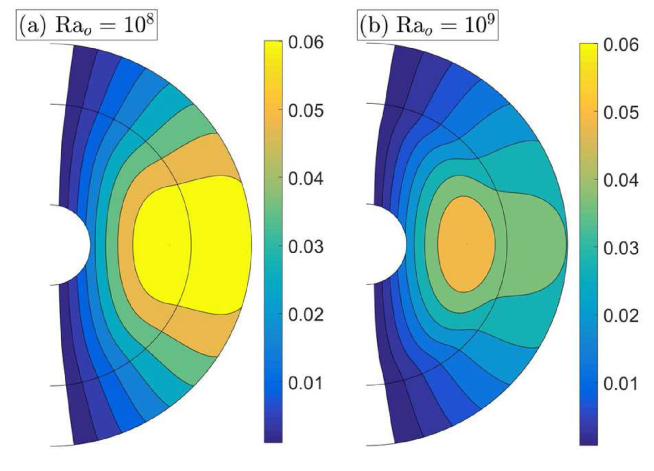} \\
    \caption{Top: 3D visualizations of the tachocline confinement scenario
      with a fossil field, adapted from \cite{2011A&A...532A..34S}. The
      convective motions are shown in the lower cut-out of the sphere,
    with orange-red volumes denoting upward motions and blue volumes
    downward motions. A few selected field lines are traced from the
    middle of the radiative zone. The initial confined dipole is shown
  on the left panel. In the middle panel, the field starts to be
  sheared in the lower tachocline. In the right panel, the field has
  permeated through the tachocline and is now connected to the
  surface. Bottom: confinement of a fossil field through convective turbulent pumping at low (left) and high (right) Rayleigh numbers, adapted from \cite{2021MNRAS.503..362K}. \FinalModif{The black contours show the dipole field lines with values of the streamfunction depicted by the colormap.}}
    \label{fig:StrugTacho}
\end{figure}

The confinement of the solar tachocline by a fossil field was proposed
by \cite{1998Natur.394..755G}. This scenario was based on equilibrium arguments,
but lacked the geometrical realism of a 3D modelling. In particular,
it relies on the existence of a meridional flow in the upper
tachocline which prevents the magnetic field to permeate
above. Conveniently, the radiative spreading theory of
\cite{1992A&A...265..106S} predicts that such a meridional circulation
necessarily co-exists as the differential rotation burrows down in the
radiative core on secular timescales \citep{2012ApJ...755...99W}. Nevertheless,
this scenario has one major difficulty (setting aside the aforementioned inner zone of tachocline assumed to be magnetic field-free). Indeed,
such a meridional flow has to go down at some
latitude and up at some others to conserve mass. In the upwelling
part, the magnetic field is likely to be entrained into the convection
zone and therefore does not remain confined. This is what was initially
found in numerical simulation of this confinement scenario in
\cite{2006A&A...457..665B,2011A&A...532A..34S} using the ASH code. 3D
visualizations of this phenomenon are shown in Figure \ref{fig:StrugTacho}. Initially (left panel), a fossil
magnetic field is assumed to be confined inside the radiative core of
the Sun. The model is then evolved, and the magnetic field is twisted
in the tachocline (as seen in the middle panel). It eventually finds a
way to connect to the convective envelope (third panel) as was just
explained. This connection path can prevent the fossil field 
scenario \citep{1998Natur.394..755G} to establish if the angular momentum exchanged by
magnetic torques between the convective and the radiative zones is strong enough. Indeed, the
first consequence is that the radiative core is not in solid body
rotation anymore, which is of course in disagreement with the
observational constraints at our disposal (see Figure
\ref{fig:DRThompson}).

The modelling efforts of \cite{2011A&A...532A..34S} were revisited by
\cite{2013MNRAS.434..720A} because their parameter regime was
preventing a true radiative burrowing of \cite{1992A&A...265..106S} to
occur, by instead favoring viscous burrowing. Indeed, the meridional flow in their case was solely driven by
the differential rotation inside the convection zone, and was
consequently less energetic and less extended than a meridional
circulation accompanying a radiative burrowing
\citep{2018ApJ...853...97W}. In this latter work, they
unambiguously showed that the differential rotation can spread
downward in various ways depending on the
adimensional parameter
\begin{equation}
  \label{eq:sigma}
  \sigma = \sqrt{\Pra}\frac{\vert N \vert}{2\Omega_\odot} < 1\, ,
\end{equation}
where $\Pra=\nu/\kappa$ is the Prandtl number. If $\sigma > 1$, the differential rotation spreads
viscously. If $\sigma < 1$, the spread is controlled by thermal
dissipation and a non-negligible meridional circulation develops as
the differential rotation burrows. This regime is particularly challenging to model in the solar
tachocline, because the Prandtl number is expected to be much less than $10^{-3}$ there. New 3D global simulations of this problem, carried out in the correct $\sigma$-parameter regime, indicate that the meridional flow can indeed halt efficiently the outward magnetic spreading in the equatorial region compared to the results of \citet{2011A&A...532A..34S}. Nevertheless, even in this regime the regions where the meridional flow turns upwards still allow the magnetic field to permeate into the overlying convective envelope and magnetic torques lead to the internal radiative zone to depart from solid-body rotation (Strugarek \& Brun, in prep.).   

Concurrently,  another mechanism may contribute to the confinement of the magnetic field in the radiative interior. The turbulent magnetic pumping is a mean-field effect that advects a large-scale field away from the regions where turbulence is strongest \citep{1998ApJ...502L.177T}. 
According to mean-field theory, the amplitude of the advection is proportional to minus the gradient of the turbulent magnetic diffusivity, $\gamma = - \frac{1}{2} \nabla \eta_{\rm T}$ \citep{2008AN....329..372K}. Thus, it is expected that the field is transported from the convection zone downward into the radiative interior. Alternatively, the upwards diffusion of the fossil field could be balanced by its turbulent advection. In mean field models, this effect was verified by \cite{2008AN....329..372K}. They found that total confinement can be found for $\nabla \eta_{\rm T} \gtrsim 10^5$. The recent global simulations of \cite{2021MNRAS.503..362K} explore this effect in the Boussinesq approximation and in the non-rotating case.  They can progressively increase $\nabla \eta_{\rm T}$ by increasing the Rayleigh (${\rm Ra}$) number in the convection zone, i.e., by making more vigorous convection. In their highest resolution case, with ${\rm Ra} =10^9$, their simulations evince confinement of the fossil field (see bottom panels in Fig.~\ref{fig:StrugTacho}). Yet, a much higher ${\rm Ra}$ is still needed to achieve the advection values predicted by  \cite{2008AN....329..372K}. So far, it is also unclear how rotation may affect these results. In addition, this mechanism relies on the strength of convection in the deep convection zone, which is debated today based on various helioseismic inversions and numerical simulations results \citep{2015SSRv..196...79H,Chapter10}. 

To conclude, the present understanding of the tachocline confinement scenario by a fossil magnetic field can be summarized as follows:
\begin{itemize}
    \item In the correct burrowing parameter regime of the solar differential rotation (i.e. the radiative spreading), the meridional flow accompanying the burrowing can efficiently halt the outward spread of the hypothetical fossil magnetic field where the flow and the magnetic field are horizontal.
    \item This spread cannot be easily halted in upflow regions of the meridional circulation, and it remains in question to know how the solar convection zone can reach the level of turbulent magnetic pumping needed to compensate the loss of confinement of the fossil field in these regions.
    \item If the spread of the fossil field cannot be halted at some specific latitudes, a magnetic connection between the convective and radiative zone establishes. The associated angular momentum transport leads to a significant departure of the inner radiative zone from solid-body rotation, that is incompatible with present constraints from helioseismology.
\end{itemize}

\section{Global (M)HD of the solar tachocline}
\label{sec:inst-solar-tach}




\FinalModif{After having reviewed the confinement mechanisms of the solar tachocline, we now turn to the more generic MHD phenomenon that could be taking place in this particular region of our Sun, and that can affect the large-scale flows and the magnetism of our star.}

Differential rotation (DR) inferred from observations in the tachocline (see Fig. \ref{fig:DRThompson}) can be 
sustained 
by the angular momentum transport in latitude and radius, due to the influence of rotation on convection. This differential rotation can itself be unstable to global hydrodynamic disturbances of low longitudinal wave number $m$. If the tachocline differential rotation is perturbed by some random disturbances or by disturbances that include some longitudinal structures in the form of normal modes, these disturbances either can grow by extracting energy from the differential rotation, or decay away by transferring energy to the differential rotation \citep{2007sota.conf..243G,2012ApJ...745..128D}. In the nonlinear regime they can quasiperiodically exchange energies with the differential rotation, and form patterns that have properties of Rossby waves. In particular, the disturbance patterns propagate in the longitudinal direction and have a tilt in latitude. 
Because of their latitudinal tilts they are able to either extract energy from the DR or can give energy back to the DR. 

In addition, a large amount of toroidal magnetic field is expected to exist at the solar tachocline due to its subadiabatic stratification and the strong radial and latitudinal shear.
Therefore, toroidal fields in the tachocline are likely closely coupled by MHD processes to the differential rotation and the aforementioned hydrodynamic disturbances, leading to global MHD instabilities. Their behavior depends on the relative energy present in the differential rotation compared to that of the toroidal field \citep{1983MNRAS.204..575M,1993ApJ...417..762C,2004IAUS..215..366C,2004IAUS..215..356S}. 
The disturbances of toroidal fields, coexisting with differential rotation, can again produce modes which retain some properties of Rossby waves, particularly their phase speed in longitude and latitudinal tilt structures \citep{1997ApJ...484..439G}. But both differential rotation and toroidal fields change these speeds significantly compared to the well known retrograde propagating Rossby waves for a uniformly rotating thin spherical shell of constant thickness, first found by \citet{1940TrAGU..21..262H}. 

Moreover, the presence of magnetic field within the stably stratified radiative interior of stars (the tachocline being subadiabatic qualifies it as such a region) further leads to a large range of MHD instabilities \citep{1999A&A...349..189S,2015A&A...575A.106J}.
Several MHD instabilities present in stellar radiative interiors have hence been studied in details, for instance that of a large scale poloidal field with closed field lines \citep{1973MNRAS.163...77M,1973MNRAS.162..339W,2007A&A...469..275B}, that of strong toroidal magnetic field (akin to the tachocline situation) \citep{1973MNRAS.161..365T,1985MNRAS.216..139P,2019MNRAS.490.4281G} and the MRI \citep{2006ApJ...650.1208M,2011MNRAS.411L..26M,  2020A&A...641A..13J} to cite only a few emblematic ones. It was for instance shown that the ratio between the poloidal and toroidal magnetic field components or energies is key to maintaining or not the magnetic structure into a stable configuration on long timescales \citep{1973MNRAS.162..339W,1980MNRAS.191..151T,2004Natur.431..819B,2008MNRAS.386.1947B,2009MNRAS.397..763B,2010A&A...517A..58D, 2023MNRAS.521.1415M,2023MNRAS.520.5573F}. Special attention has been given to the interaction between a large scale field with a large scale mean flow (meridional circulation or differential rotation, \citealt{1977MNRAS.178...27M,1987MNRAS.226..123M}). In the situation where a large scale shear is present and maintained, it was shown that a dynamo loop could exist, using the shear as $\Omega$-effect and the generation of non axisymmetric fields and flows through MHD instabilities as $\alpha$-effect \citep{2002A&A...381..923S,2003ApJ...599.1449C,2006A&A...449..451B,2007A&A...474..145Z,2022MNRAS.517.3392Z,2023Sci...379..300P}. However, such dynamo loop in the radiative interior, with the large scale shear being maintained by external processes such as the convective envelope above or the angular momentum extraction due to stellar wind torque, are not easy to realize as demonstrated in \cite{2007A&A...474..145Z}. Indeed, they required an effective $\alpha$ effect originating from the correlations of the field and flow perturbations, i.e. corresponding to strong electromotive force $\epsilon = \langle{ v' \times b' \rangle}$. Such correlation can be provided by the MHD instabilities but are not easy to achieve or maintain on long timescales because they can easily be damped when the large scale field aligns with the large scale shear flow or vice and versa, a phenomenon known as Ferraro's law of iso-rotation. Nevertheless, it is remarkable that the joint presence of magnetic fields (either poloidal, toroidal, or both) and large scale shears yields a rich realm of physical mechanisms that are likely acting within stars such as our Sun and in their tachoclines.
Since both HD and MHD global processes are likely to operate in the tachocline, we discuss them in more details in the following subsections.

\subsection{Global HD instabilities}
\label{sec:GlobalHDinst}

In strictly 2D spherical shells, a DR profile 
of the solar type\footnote{such as expressed by $\Omega= \Omega_0 - s\, \sin^2\theta$ form, in which $\Omega_0$ denotes the core-rotation rate (or, in other words the rotation rate at $32^{\circ}$ latitude) and $s$ the amplitude of equator-to-pole differential rotation.} is unstable to low longitudinal wave numbers only if the equator-to-pole difference in rotation is nearly $30\%$ \citep{1981GApFD..16..285W}. Nevertheless, the instability can occur for smaller equator-to-pole contrast if (i) $\Omega$ is expressed using both the $\sin^2(\theta)$ and $\sin^4(\theta)$ terms, such as $\Omega=\Omega_0 -s_1\,\sin^2\theta -s_2\,\sin^4\theta$, and/or (ii) \FinalModif{3D effects are considered (e.g. the threshold for the DR instability drops substantially in a shallow-water model when moving from purely 2D to quasi-3D, see \citealt{1987AcA....37..313D,1999ApJ...526..523C,2001MNRAS.324...68G} )}
In the case (i), the tachocline differential rotation can be still unstable in 2D. In the case (ii), particularly for a layer like the convective overshoot layer of the tachocline that is only slightly subadiabatic on average (thus allowing more vertical motion), differential rotation can be unstable for much smaller amplitude than 30\% equator-to-pole differential rotation \citep{2001ApJ...551..536D}. In this latter case, there also appear bulges and depressions in the top surface of the tachocline (corresponding to the tachocline--convection zone interface). Because Coriolis and pressure gradient forces almost balance (as in a geostrophic balance), the perturbation flow becomes clockwise on the bulges, and counterclockwise in the depressions. These also are properties of Rossby waves patterns, whether stable or unstable. 

Vertical motions are also correlated with the bulges and depressions, leading to non-zero kinetic helicity. This in turn provides an additional source of alpha effect for solar dynamo models \citep{2001ApJ...559..428D} that \FinalModif{could play a role} in parity selection in solar dynamo models \citep{2002A&A...390..673B,2015ApJ...806..169B}.

\subsection{Rossby waves}

In the Earth's atmosphere, where Rossby waves were first discovered, the forcing primarily comes from differential heating from the Sun; this is thermal forcing. Coriolis forces are always present in a rotating body, but they do not provide a source of energy to drive the Rossby waves. They only alter the direction of fluid motion in the wave. In the case of the Sun, unlike the Earth's atmosphere, the energy is coming from the mechanical forcing of differential rotation at the top of the tachocline. However, like in the Earth's atmosphere, the Coriolis forces again can alter the direction of plasma flow, but can't provide energy. 

Very much like the Earth's Rossby waves, solar Rossby waves were also first predicted theoretically before they were directly observed \citep{1987AcA....37..313D,1997ApJ...484..439G,2010ApJ...724L..95Z}. These works found that the disturbances that perturb the tachocline differential rotation form the patterns that have Rossby waves-like properties such as the ones described in Section \ref{sec:GlobalHDinst}. Later on, various observations have indicated the existence of Rossby waves in the Sun \citep{2017NatAs...1E..86M,2018NatAs...2..568L,2022ApJ...931...54H}. 

\FinalModif{In fact, even before the role of Rossby waves as perturbation to the global flows and magnetic fields, the idea that the Sun might have an MHD dynamo driven by Rossby waves was explored by \citet{1968AJS....73R..61G,1969SoPh....8..316G,1969SoPh....9....3G}. That model assumed the existence of a 'thermal wind' inside the convection zone, with latitudinal and radial rotation gradients as well as an associated latitudinal specific entropy gradient. Gilman found that Rossby waves and DR can together drive a cyclic dynamo with a period within an order of magnitude of the observed sunspot cycle. The solar tachocline, for which the model assumptions were plausible, was not discovered until almost thirty years later, and observational evidence of Rossby waves in the Sun has been found only over the past five years.}

So-called thermal Rossby waves have 
been discovered theoretically for both the convection zones and radiative interiors \citep{1981ApJS...45..381G,2022ApJ...932...68H}. Those may be more like Earth's Rossby waves, in that they contain thermodynamic effects and rely for some of their properties on the outward decline in plasma density, as well as on the spherical geometry of a thick layer as opposed to thin layer
that is the tachocline. Thermal Rossby waves are yet to be detected in the Sun.

\FinalModif{It is worth noting that i}n the presence of a latitudinal DR that is stable to perturbations, Rossby waves can \FinalModif{theoretically} still exist. \FinalModif{In that case, t}heir patterns do not have tilts in latitude, and so they cannot extract energy from the reference state. Their phase velocity differs from the uniform rotation case by an amount close to that of the local rotation speed at the latitude where the relative amplitude of the Rossby wave peaks. 
In an unstable differential rotation, 
Rossby waves behave similarly, but they have tilts in latitude and a phase speed in longitude close to the rotation speed of the latitude where the unstable mode peaks. These properties are similar for purely 2D waves, and the quasi-3D waves characteristic of hydrodynamic shallow water models \citep{2001ApJ...551..536D}.

With the addition of a toroidal field, the picture changes significantly, depending on the latitudinal profile and strength of the field.
\FinalModif{We first note that} Rossby waves are still nearly hydrodynamic (with retrograde propagation). In addition, a very slow prograde wave starts to appear, and vanishes in the limit of zero toroidal field \citep{2020ApJ...896..141D,2020JFM...904R...3H}. For much stronger fields, magnetic effects overpower Coriolis forces, leading to Alfvén waves propagating with nearly equal speeds in both prograde and retrograde directions. Unstable MHD waves for even moderately strong but latitudinally compact toroidal fields propagate at the rotational speed of the latitude  where the toroidal field peaks. Therefore, these waves are likely to be retrograde at high latitudes (poleward of sunspot zones), but could be prograde for a certain range in low latitudes \citep[see, e.g.][]{2018ApJ...862..159D}.

It is to be noted that all Rossby waves are inertial waves, however all inertial waves are not Rossby waves. Recently, solar inertial oscillations were studied by \citet{2022ApJ...934L...4T} in a 3D spherical shell model of the convection zone containing a homogeneous, incompressible, viscous fluid, using an eigen-system formalism. Retrograde as well as prograde Rossby modes were found, along with other inertial modes, which are distinct from Rossby modes and have radial velocities comparable to horizontal velocities. Rossby modes found by \citet{2022ApJ...934L...4T} are similar to those found by \citet{2022A&A...662A..16B} using a similar model but in a fully compressible regime.

\subsection{Global MHD instabilities}

The addition of a toroidal field to 2D, quasi-3D shallow-water type, and 3D thin-shell type hydrodynamical models 
has a destabilizing
effect, \FinalModif{as we noted in the previous section}. In that case, instabilities for low longitudinal wave numbers can occur down to very small differential rotations. Which longitudinal wave number is most unstable depends on the strength and profile of the toroidal field. Narrower toroidal bands render higher longitudinal wavenumbers unstable. But very narrow bands of about $2^{\circ}$ latitudinal width are not unstable because they cease to sense the difference in rotation rate across the band. So there is an optimum bandwidth for instability \FinalModif{in the context of the solar DR profile} \citep{2000ApJ...528..552G}.

The existence of unstable toroidal bands at particular latitudes with low longitudinal wave number could provide templates for the longitude and latitude distribution of active regions on the solar surface. The bulges in the tachocline that happen to contain strong toroidal field may indeed be the source for emerging magnetic flux that produces new active regions \citep{2017NatSR...714750D}.


Global HD and MHD instabilities in the tachocline evolve non-linearly due to the interactions among differential rotation, magnetic fields and Rossby waves, which are essentially the disturbance patterns. Nonlinear evolution of these instabilities can produce many interesting features which are discussed in the following two subsections, for pure 2D models and quasi-3D thin-shell shallow-water type models.

\subsubsection{Nonlinear evolution of 2D HD and MHD instabilities}

In a pure 2D tachocline, nonlinear evolution of the HD instabilities can produce high-latitude jets. Due to the action of Reynolds stresses, the angular momentum can be extracted from the differential rotation specifically from the mid-latitudes, and can be deposited on the poleward sides, creating prograde jets there. Traces of such tachocline jets were observationally detected by helioseismic analysis \citep{2005ASPC..346..115C}. 

For a 2D magnetized tachocline, nonlinear evolution of MHD instabilities lead to two interesting features. \FinalModif{In the case of} broad toroidal fields open into clam-shell pattern \citep{2001SoPh..199..231C}. \FinalModif{Conversely,} a narrow toroidal band produces tipping or deformation patterns depending on the field strength. If the toroidal band \FinalModif{harbours strong fields}, it behaves like a 'steel' ring, and it tips to produce an $m=1$ longitudinal pattern \citep{2003ApJ...582.1190C}. \FinalModif{If the field is weak,} the ring deforms and produces $m>1$ patterns in longitude.

\subsubsection{Nonlinear evolution of quasi-3D instabilities}
\label{sec:NLE_3D}

In quasi-3D thin-shell and shallow-water type models, 
the formation of high-latitude jets, 
as well as the clamshell opening of broad toroidal fields and tipping of narrow bands occur too. \FinalModif{Nevertheless,} additional important nonlinear effects have been found in quasi-3D model simulations, which are not present in 2D models. Due to the presence of potential energy, along with kinetic and magnetic energies in a shallow-water model, nonlinear interactions among differential rotation, toroidal field and Rossby waves can produce the 'Tachocline Nonlinear Oscillations' (TNOs), which arise as a consequence of quasi-periodic exchange of energies among \FinalModif{these three reservoirs}. These oscillations are produced in both HD and MHD cases, very much like 'nonlinear Orr mechanism' in fluid mechanics. The period of oscillation is typically 6-18 months \FinalModif{in the context of the solar interior}. \FinalModif{This oscillation period interestingly coincides with} the observations of Rieger-type periodicity, or 'seasons' of solar activity, characterized by intervals of more intense activity alternating with more quiet periods. \FinalModif{This suggests that the origin of such 'seasons' of could be the TNOs} \citep{2017NatSR...714750D}. 

\begin{figure}[h]
    \centering
    \includegraphics[width=\linewidth]{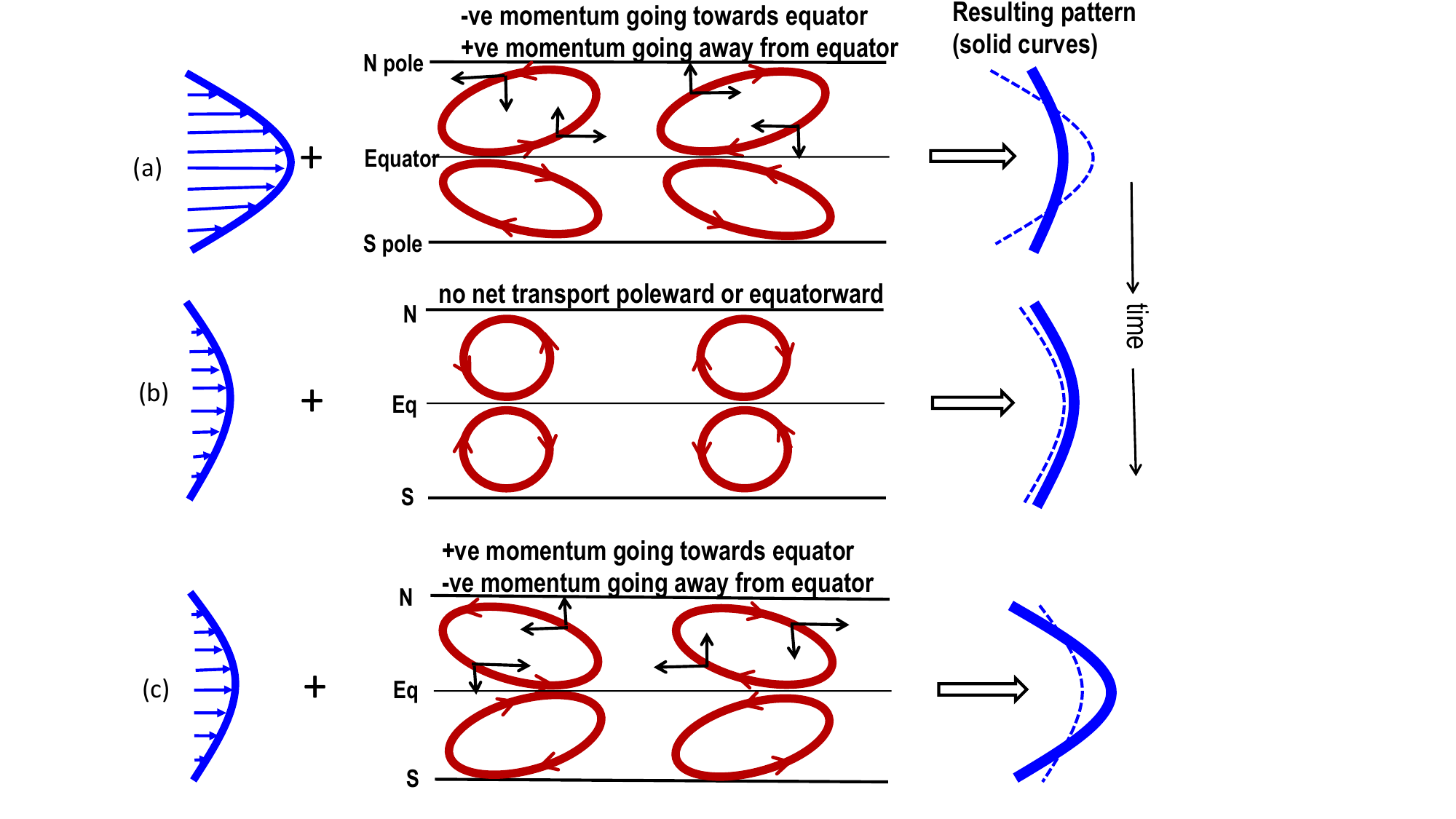}
    \caption{Panels a, b, c show schematically how Rossby waves and tachocline latitudinal differential rotation nonlinearly interact. Time goes from top panel to bottom. Energy from unperturbed differential rotation (thick blue curve in left frame of panel(a)) is extracted by perturbation flow-patterns (eastward-oriented red ellipses in the north hemisphere, plotted in middle frame of panel (a)), because their eastward tilts transport angular momentum away from the equator; consequently the polar region spins up and the equatorial region spins down, resulting into a decrease in pole-to-equator differential rotation (thick blue curve in right frame of panel a), compared to the original (thin dashed curve). Panel (b) shows that tilts become neutral when no more energy can be extracted out of differential rotation. However, the perturbation flow-patterns overshoot from neutral tilt to acquire westward tilts, as shown in panel (c) (middle frame); hence flow-patterns can transport positive angular momentum back from high to low latitudes (see the black arrows on the flow-patterns of the bottom panel), spinning up the equatorial region again. Consequently, differential rotation is restored back, and becomes the source of energy to be extracted by the perturbation flow-patterns. In a nearly dissipationless system this process repeats, very much like nonlinear Orr mechanism in fluid dynamics, and leads to an oscillatory exchange of energies between Rossby waves and tachocline differential rotation \citep{2017NatSR...714750D}.}
    \label{fig:TNO-Physics}
\end{figure}

Figure \ref{fig:TNO-Physics} describes the physics of TNOs in an ideal hydrodynamic system, in which they occur due to nonlinear exchange between kinetic energy of differential rotation and perturbation energy of Rossby waves. 
\FinalModif{We note that the TNO physics in the MHD case is a bit more complex in the MHD case, and defer the interested reader to Figure 6 of \citet{2018ApJ...853..144D} for their extended description}. \FinalModif{An interesting property of TNOs is that they} do not damp out in several years' runs in an ideal MHD simulation. \FinalModif{Evidence for their existence has been furthermore obtained recently} in a fully nonlinear direct numerical simulation of tachocline instabilities using generalized quasilinear approximation by \citet{2019JPlPh..85a9013P}, \FinalModif{where the} inclusion of magnetic diffusivity \FinalModif{is shown to} create an abrupt transition of short-term to long-term oscillations between kinetic and magnetic energies. 

\begin{figure}
\centering
    \includegraphics[width=.4\linewidth]{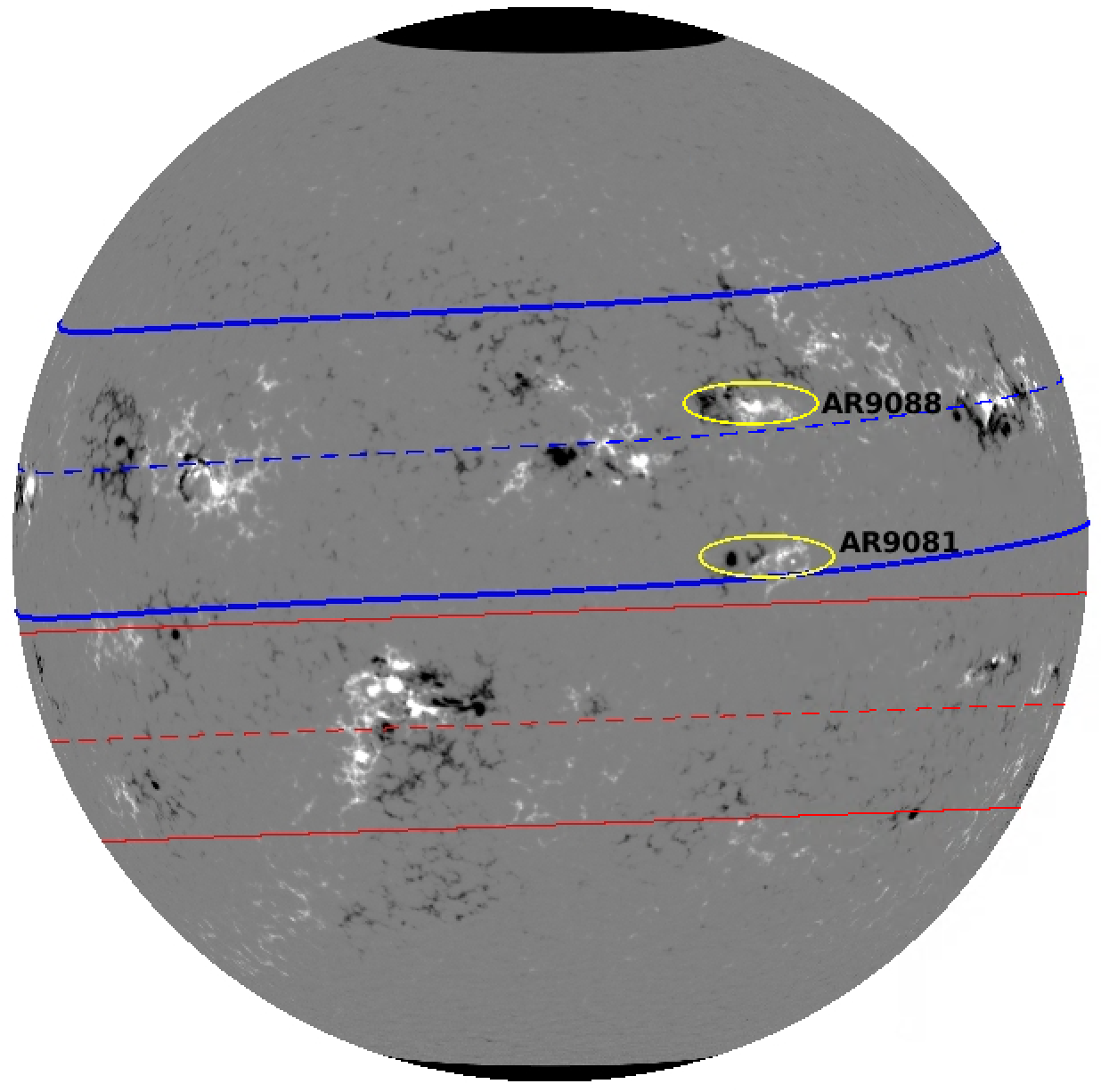}
    \includegraphics[width=.4\linewidth]{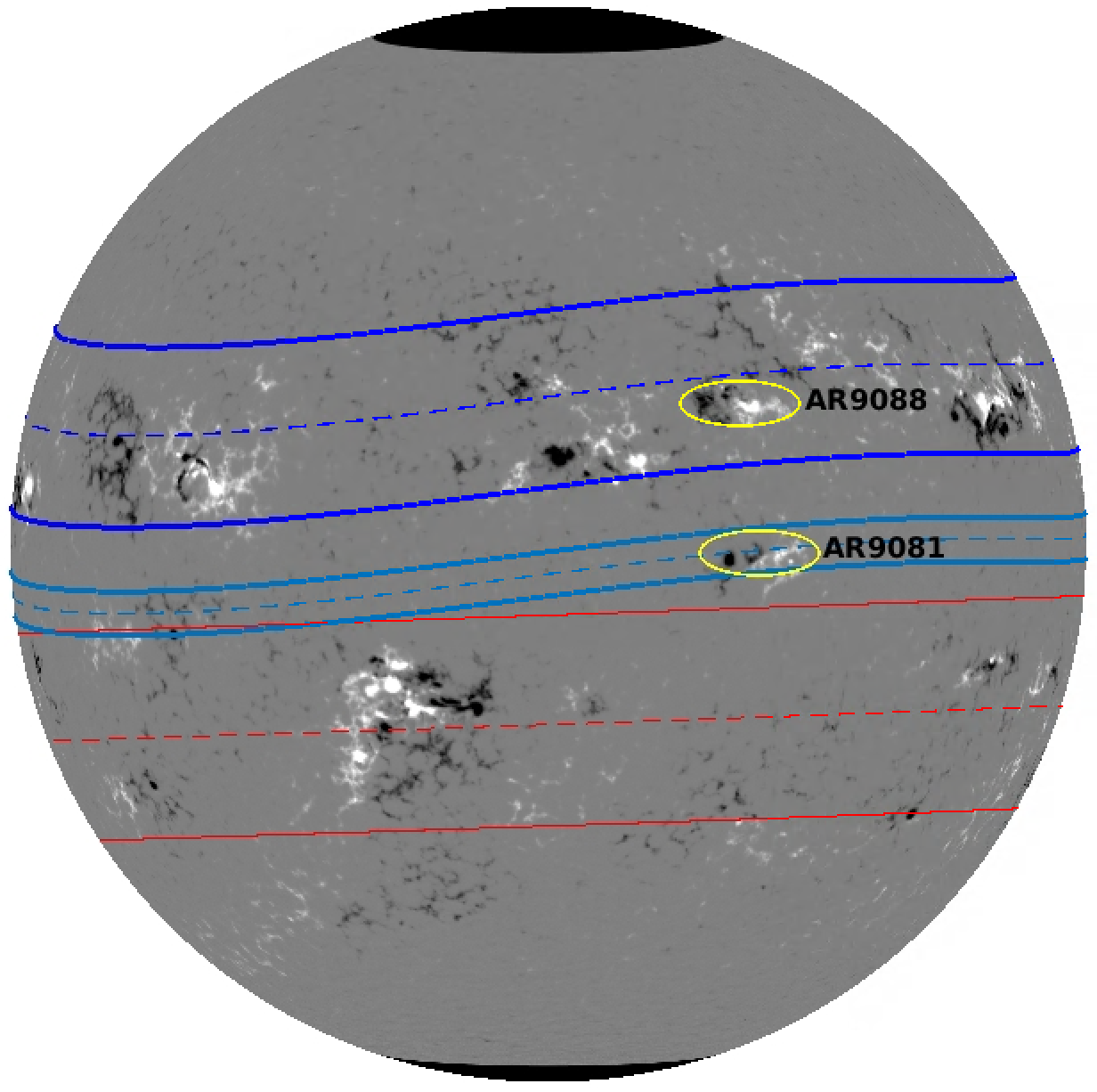}
  \caption{Left panel displays active regions' toroid-pattern
for July 19, 2000, if all the active regions emerged from one wide toroidal magnetic band
from the base of the convection zone; the right panel displays two split
toroid-patterns if active regions emerged from spot-producing toroidal rings
that have undergone dynamical splitting. North and South toroids are
displayed in blue and red respectively \citep{2021ApJ...922...46D}.}
  \label{fig:spots-from-split-rings}
\end{figure}

Another notable nonlinear effect arises from the action of the magnetic stress (often called the 'mixed stress'), which arises due to the cross-correlation between perturbation magnetic and velocities fields in unstable global MHD modes. This stress can cause extraction of energy from the center of the band, and deposition of energy on the both shoulders of the band. In the nonlinear evolution, this process can continue, and at a point the energy at the band-center can be depleted so much as to cause the band to split into two. The poleward part of the split-ring slips fast towards the pole due to curvature stress, which is responsible for causing the famous "poleward-slip" instability of a toroidal ring \citep{1982A&A...106...58S}. A $6^{\circ}$ toroidal ring splitting into two can be seen in this ring-split \href{movie}{https://drive.google.com/file/d/1RUicqKRgnEmiFB9EGU0XpZSSoT-vBTw5/view?usp=sharing}. The toroidal ring is displayed in white arrow-vectors on the colormap, which represents deformation of thin fluid-layer's top-surface (red/orange denotes bulges and
blue/dark-blue depressions). The movie shows that the ring is just split into two at t=6.7023 (24.5 days), and the split part of the ring is seen to move poleward at t=7.6556 (30
days).
The snapshots are presented in dimensionless time (dimensional time can
be obtained by multiplying by 3.65 in the unit of days). 

The equatorward part can move slowly to the equator because of the energy getting deposited from the band-center to the equatorward shoulder of the band. This provides a possible physical mechanism for a pair of spots, often emerging at the same longitude but separated in latitude by more than 20-degrees \citep{2021ApJ...922...46D}. The Figure \ref{fig:spots-from-split-rings} shows an observational evidence of a pair of spots that could have emerged from split-ring.

\begin{figure}
\centering
    \includegraphics[width=\linewidth]{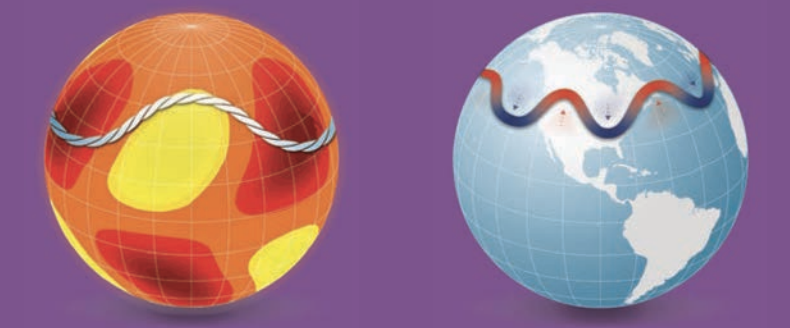}
  \caption{ \FinalModif{Large-scale Rossby waves propagating around the Sun may play a role in triggering solar storms (yellow and
deep red regions indicate where sunspot activity is unlikely and more likely, respectively). These waves are akin
to Rossby waves in Earth’s atmosphere, which influence weather, including potential heat waves (red arrows)
and cold snaps (blue arrows) (schematic figure adapted from SCIENCE NEWS BY AGU, Eos.org 51, vol. 101, no. 8, August 2020.)}}
  \label{fig:solar_RW_schem}
\end{figure}

MHD Rossby waves \FinalModif{could have an effect} on the solar magnetic activity cycle \citep{2015ApJ...799...78R} 
and on space weather \citep{2020SpWea..1802109D}.  \FinalModif{Here for the general readers, we display a diagram to show how magnetically modified solar Rossby waves can create spatio-temporal patterns on dynamo-generated magnetic fields at the tachocline, very much like the "jet streams" produced by the interaction of Earths' atmospheric Rossby waves with the mean zonal flow. As the "jet streams" steer the terrestrial weather by causing cold wave or draught, respectively transporting cold from high to mid-latitudes or heat from low to high latitudes, similarly spatio-temporal magnetic patterns created by the interaction of solar MHD Rossby waves could determine the plausible latitude-longitude locations of flux emergence on the solar surface (see, e.g. Figure \ref{fig:solar_RW_schem}).} \FinalModif{We point the reader to the comprehensive review by \citet{2021SSRv..217...15Z} for} more details about Rossby waves and their roles in producing various astrophysical
phenomena.

\subsection{Magnetic buoyancy instability}

\FinalModif{
Concomitantly to global HD and MHD instabilities,  perturbations to a tube or layer of magnetic field stored at the tachocline may trigger the buoyancy instability. Over the years, it has been assumed that this instability is responsible for the emergence of magnetic flux from the tachocline to the solar surface, leading perhaps for the formation of sunspots. Furthermore, the turbulence generated by this process may in principle generate a net electromotive force, and therefore an $\alpha$-effect.}

\FinalModif{The most relevant aspect of the buoyancy instability is that the conditions for its excitation are easily fulfilled for the magnetic fields developed by the strong radial shear at tachocline levels. The general idea was proposed by \cite{1955ApJ...121..491P} considering a magnetized tube of plasma in pressure equilibrium with its surroundings.  If the temperature of the plasma inside the tube is the same as in the exterior, the density inside will be smaller and the tube will become a bubble.  This out of equilibrium situation was ever since extensively studied formally as an instability.  For instance, \cite{1961PhFl....4..391N} used the energy principle of \cite{1958RSPSA.244...17B} to find the stability condition for interchange modes where the wavenumber along the field tube tends to zero (i.e., there is no deformation of the tube along the direction of the magnetic field). The instability turned out to depend on the vertical gradient of the magnetic field \citep{1975PhFl...18..490T}.  The use of the energy principle for an unstable parcel of gas was also performed for non-ideal plasmas \citep{1978JFM....85..743A}.  Yet this approach becomes insufficient for more realistic models including rotation and/or shear \citep{2004ApJ...603..785T}. These aspects, however, were studied through local instability analysis for interchange and, three-dimensional, undular modes \citep[e.g.,][]{1970ApJ...162.1019G, 1978JFM....85..743A, 1985GApFD..32..273H, 1985GApFD..34...99H}.   In general, small magnetic diffusivity and large thermal conductivity have a destabilizing effect (since the instability may develop faster than the magnetic field diffusion, and the positive entropy gradient of a stable layer may decrease, respectively). Rotation was found to have a stabilizing effect, yet the instability depends strongly on the vertical profile of the magnetic field, and rotation-induced shear may therefore contribute to make axisymmetric the easily excited non-axisymmetric buoyant modes. A complete review can be found in \citep{2007sota.conf..275H}. More recently, \cite{2018ApJ...853...65G, 2018ApJ...867...45G} extended the linear stability analysis to the spherical domain, including rotation and vertical shear, and exploring together rotational and buoyancy instabilities.  His results indicate that the subadiabatic tachocline is unstable to negative vertical shear, such as it is observed at high latitudes in the Sun. Therefore, the magnetic field diffuses rapidly there and the buoyancy instability becomes inefficient.  On the other hand, positive shear, as in the solar lower latitudes, is a stable configuration to rotational instabilities, and the buoyancy instability develops for magnetic fields above $\sim 9$ kG.  Above this threshold, plenty of unstable modes may become unstable, depending on the amplitude of the magnetic field.}

\FinalModif{The non-linear evolution of buoyant magnetic fields has been extensively studied through MHD numerical simulations.  Initial studies were performed in two-dimensional domains and explored the evolution of single flux tubes \citep{1979A&A....71...79S, 1996ApJ...472L..53M, 1998ApJ...493..480F}, or the development of interchange modes in a magnetic layer \citep{1988JFM...196..323C}.  These studies found the development of the mushroom or umbrella shape in the emerging structures. These structures develop strong vorticity at their edges which ends destroying their coherence. \cite{1990MNRAS.247P...6C} found in 2D simulations that the addition of twist to the initial magnetic field prevents the development of vorticity and facilitates the emergence of the field. It turned out, however, that in the three-dimensional case the axisymmetric perturbations are unstable only during the first stages of evolution \citep{1995ApJ...448..938M, 2000MNRAS.318..501W, 2001ApJ...546..509F}, or remain stable \citep{2001ApJ...546..509F}, depending on the ambient stratification.  The undular 3D modes are dominant, and the bending of the magnetic field lines helps the magnetic structures to remain cohesive during the emergence \citep{2001ApJ...546.1194A}.}  

\FinalModif{Concurrently, the buoyancy instability has been studied under the assumption that below the convection zone the magnetic field is organized in the form of thin flux tubes with radius much smaller than the local pressure scale height. The properties of such tubes change along them but not over their cross section. Thus, the evolution equations may be reduced to one dimension \citep{1981A&A....98..155S}. The simulations commonly consider the full longitudinal domain, and impose non-axisymmetric perturbations, which, after the instability develops, forms arch structures anchored at the base of the tube in the solar convection zone \citep[e.g.,][]{1989SoPh..123..217C,1993GApFD..72..209F} or the subadiabatic tachocline \citep[e.g.,][]{1995ApJ...441..886C}.  Because computing the evolution of thin flux tubes is generally inexpensive, a large amount of work has been performed under this approximation, exploring the role of rotation or the conditions for reproducing the tilt angle observed in active regions \citep[e.g., ][]{1987ApJ...316..788C,1993A&A...272..621D}. In order to overcome the effect of the Coriolis force, and to emerge with the tilt angles compatible with observations,  the initial field strength of the thin flux tubes must be $\ge 10^5$ G. A complete description of results and references can be found in \cite{2021LRSP...18....5F}. The most recent work on this topic \citep{2011ApJ...741...11W,2013SoPh..287..239W,2015SoPh..290.1295W} explored the evolution of thin flux tubes in the presence of turbulent convection provided by a global simulation \citep{2006ApJ...641..618M}. The results show an interesting interplay between buoyancy and advection, with initial flux tubes with strength between $40-50$ kG being the ones that better resemble observational tilt angles, and with rapid rising times. It is worth noting that the thin flux tube approximation is not except of criticism. Different works have shown that the internal dynamics of the tube,  neglected on the averaged formulation of the thin flux tube approximation,  may play a relevant role during the emerging process (\citealt{1998MNRAS.298..433H,1998ApJ...509L..57H}, or more recently \citealt{2015ApJ...814....2M}).}  

\begin{figure}
\centering
    \includegraphics[width=1.0\columnwidth]{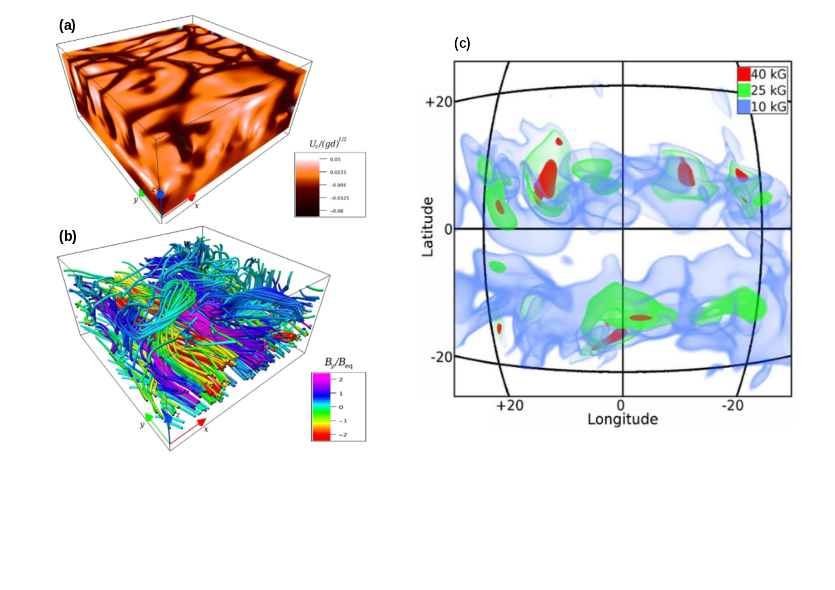}
  \caption{ \FinalModif{Panels (a) and (b): convection pattern and magnetic field lines resulting from a numerical simulation with forced radial shear beneath a convectively unstable layer. If the shear is sufficiently strong, the magnetic elements become buoyant forming $\Omega$-loops. This magnetic field modifies the pattern of convection \citep[adapted from][]{2011A&A...533A..40G}. Panel (c): volume rendering of the buoyant magnetic elements developed from a global dynamo simulation. The iso-surface in blue is in equipartition with the kinetic energy whereas the regions in red have four times more energy \citep[adapted from][]{2013ApJ...762...73N}. }}
  \label{fig:buoyancy_figs}
\end{figure}

\FinalModif{In the same spirit, Cartesian three-dimensional MHD simulations have explored the evolution of imposed magnetic tubes in the presence of convection \cite[of course this not exactly happening at the stable tachocline but above it, see][]{2003ApJ...582.1206F}, the self generation of a magnetic layer by a vertical shear \citep{2008ApJ...686..709V, 2009ApJ...690..783V}, or even the vertical shear beneath a convectively unstable layer \citep{2011A&A...533A..40G}. Convection strongly affects the evolution of a magnetic structure and creates arched tubes anchored by the downflow lanes with a fat magnetic flux configuration frozen-in with the convectively rising plasma.  Under shear conditions more extreme that in the solar tachocline, undular magnetic structures develop and may become buoyancy unstable (Fig.~\ref{fig:buoyancy_figs}(a) and (b)). If the shear is sustained by a mechanical forcing then a self-sustained dynamo can exist because an incoherent, yet different from zero, $\alpha$-effect can persist \citep{2011A&A...533A..40G}.  An oscillatory behavior can also be found if the field is replenished by the boundary condition \citep{2007ApJ...663L.113K}.   }

\FinalModif{Three dimensional simulations have also been performed in spherical coordinates considering the emergence of magnetic flux tubes in adiabatic \citep[e.g.,]{2007AN....328.1104J,2008ApJ...676..680F,2017A&A...607A...1F} and superadiabatic layers \citep{2009ApJ...701.1300J}.  Even though these works preclude the development of the buoyancy instability in the subadiabatic tachocline, they clearly expose how complex it is for a buoyant structures to cross a strongly stratified layer, specially in the presence of turbulent convection and its associated large scale mean flows \citep{2009ApJ...701.1300J} or of background dynamo magnetic fields \citep{2013ApJ...772...55P}.  Reproducing the observed features of active regions, therefore, requires a fine tuning of parameters as the magnitude of the magnetic field, twist, and wave-number of the instability.  Note that the origin and magnitude of the twist considered in these models do not have proper justification.  Most recently \cite{2018ApJ...859L..27M,2021ApJ...909...72M} explored the evolution of  twisted magnetic tubes embedded in an adiabatic layer with a background magnetic field. They found that there is a lower limit of the background field that allows the tube emergence, and that magnetic structures where the twist and the background field are aligned rise more easily than tubes with the opposite configuration.   Finally, magnetic buoyant elements have been observed in convective global dynamo simulations where the magnetic field is not imposed but develops self-consistently \citep{2011ApJ...739L..38N,2014SoPh..289..441N,2014ApJ...789...35F}.  These works showed that convection may not be a hindrance to magnetic flux emergence but may contribute to the rise of magnetic loops (see Fig.~\ref{fig:buoyancy_figs}(c)). It is not yet clear, however,  whether these buoyant magnetic elements may reach the surface levels and form bipolar regions. Another kind of simulations have been performed to explore the conditions of the surface emergence phenomena \cite[e.g., see the review of][and references therein]{2014LRSP...11....3C}. These models are far from the scope of this review.}

\subsection{Double band and the extended solar cycle}

The examination of torsional oscillations (\citealt{2016ApJ...828L...3G}), the 
coronal green line emission (\citealt{2013SoPh..282..249T}) and the studies of 
ephemeral active regions (\citealt{2018FrASS...5...17M}) provide hints of the
existence of extended solar cycles. In these observables, the solar activity cycles also
manifest at higher 
latitudes (around 60-70 degrees) than the classical butterfly diagram for the sunspot cycle. It is generally observed at these latitudes ahead of the start of 
the new sunspot cycle. 
This high-latitude component of the solar cycle gradually moves towards the equator, maintaining a separation of about $30^{\circ}$ (see, e.g., the figure 1 of \citealt{2023ApJ...945...32B})
with the low-latitude active cycle's branch. When the active cycle's branch reaches near the equator to be ready to annihilate with its opposite-hemisphere counterparts, the high-latitude branch reaches about 30-35 degrees, 
and could then start producing sunspots of the next cycle.
Motivated by these observations, one may consider the possibility that there are double, subsurface 
magnetic bands with opposite directions in each hemisphere, with a low-latitude 
toroidal strip and a weak high-latitude strip. 
\citet{2023ApJ...945...32B} studied the global HD/MHD instabilities of such a double structure to understand
the global MHD processes and their contribution to properties of solar activity 
and the solar cycle. The thin solar tachocline can be modelled in a shallow-water formalism
as a 3D thin fluid 
shell with a rigid bottom and deformable top surface. The flows and fields are much 
larger in the latitude-longitude than in the radial direction. This condition is 
validated in the tachocline because of its subadiabatic stratification.

\begin{figure}
\centering
    \includegraphics[width=1.0\columnwidth]{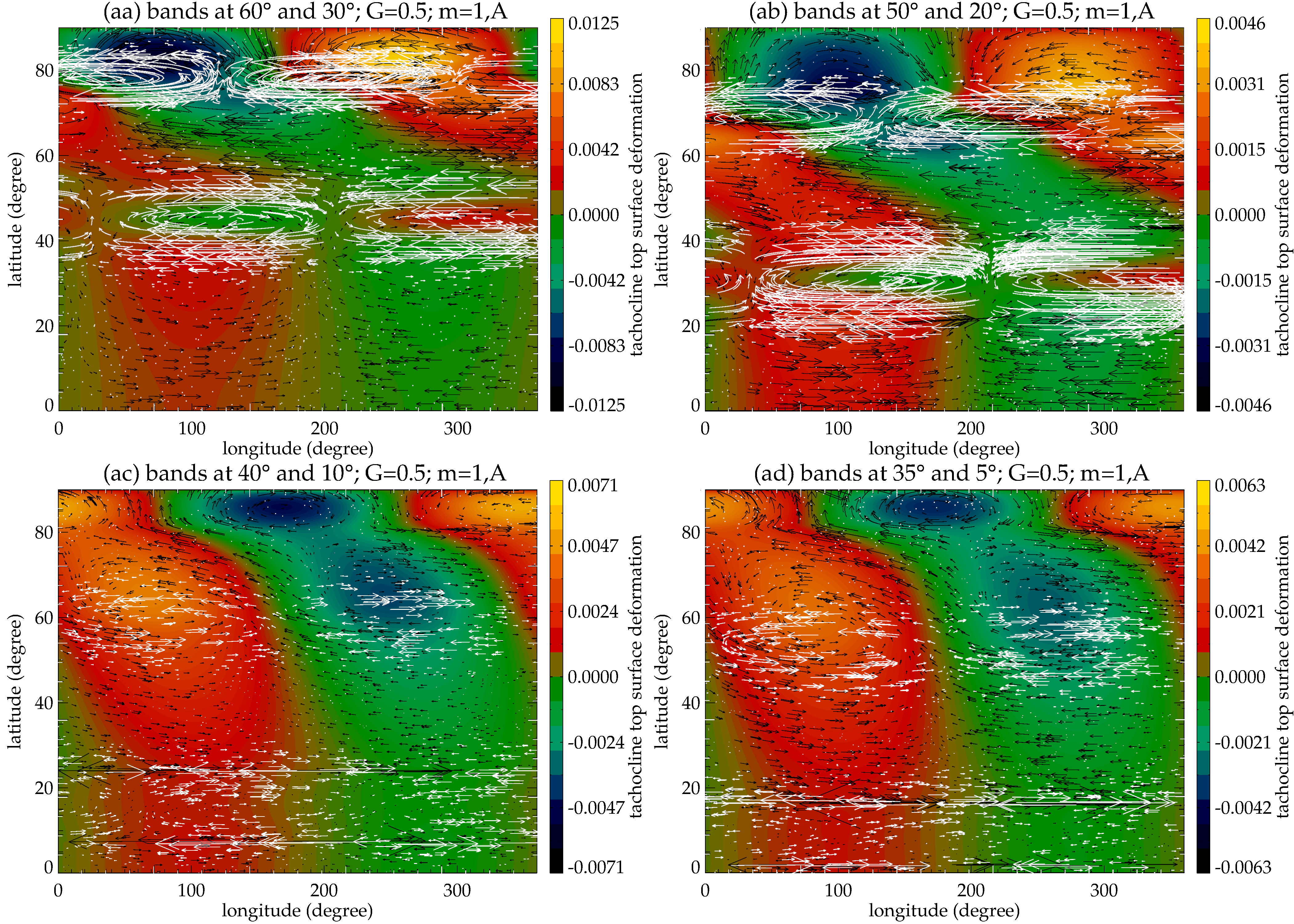}
  \caption{Snapshots of velocity and magnetic disturbance patterns are shown in the latitude-longitude plane for the asymmetric {\it m}=1 modes. Black arrows represent magnetic field vector, white arrows flow vectors and color map the tachocline top surface deformation (red-yellow denote the bulging of top surface, and green-blue the depression). This figure shows the eigenfunctions for the overshoot part of tachocline ($G=0.5$) and the bands at $60^{\circ}-30^{\circ}$ (aa), $50^{\circ}-20^{\circ}$ (ab), $40^{\circ}-10^{\circ}$ (ac) and $35^{\circ}-5^{\circ}$ (ad) \citep{2023ApJ...945...32B}.}
  \label{fig:doubleband}
\end{figure}

Motivated by observations of extended solar cycle, \citet{2023ApJ...945...32B} considered a double-band system \FinalModif{located within the solar tachocline and} consisting of two oppositely-directed toroidal bands in each hemisphere, and study their MHD instability in a quasi-3D MHD shallow-water model. 
As we make the two oppositely-directed bands in each hemisphere migrate equatorward in the model \citep{2023ApJ...945...32B}, it is found that
the high- and low-latitude bands 
interact in the same hemisphere when they are at latitude $60^{\circ}$ and 
$30^{\circ}$ (Figure \ref{fig:doubleband}).  
Here, band-interaction occurs by teleconnection mechanism -- we briefly describe the physics below, before elaborating on the MHD of the double-band in the solar tachocline.

\FinalModif{Here, band-interaction occurs by a 'teleconnection' mechanism. While teleconnection has long been known in meteorology \citep{1951JAtS....8...52L} and atmospheric circulation \citep{1984JAtS...41..961B}, this concept is relatively new for the Sun (see, e.g., \citealt{2021E&SS....801223L}). Teleconnection is essentially a contemporaneous correlation among various global fluid properties and parameters at remotely separated regions and their influence on each other. Teleconnection was demonstrated in a pioneering paper by \citet{1981MWRv..109..784W} to be created by planetary waves in the North Atlantic and North Pacific Oscillations. In the Sun, Rossby waves act as perturbations to the unperturbed DR and magnetic fields, nonlinearly exchanging energy among the perturbations and reference states of the system. This long range communication can happen in latitude (even across the equator) as well as in longitude, from one side of the Sun to the other. In the case of dynamo-generated magnetic bands in a solar interior, the basic physics of the global connections was described in figure 4 of \citet{2003ApJ...582.1190C}. }


The main conclusion from the MHD of double-band 
\citep{2023ApJ...945...32B} is that, if an extended solar cycle is represented by the two magnetic bands in the interior, generated by dynamo, namely a magnetic band at sunspot latitudes to represent 
an active-cycle 
and an oppositely-directed magnetic band at high-latitude band at 60-degree latitude to represent the extended part of the cycle, the model 
shows that the high- and low-latitude bands interact and telecommunicate among 
themselves, but not with their opposite-hemisphere counterparts, until this 
double-band-system has migrated enough towards the equator. The cross-equatorial 
communication starts when the low-latitude bands in the north and south 
hemispheres are closer than 15-degree from the equator, because then the low-latitude bands reach the optimal distance between them to start the teleconnection. This is approximately the
phase of the solar cycle when the cycle activity reaches the peak. The 
low-latitude bands in the North and South start the teleconnection across the equator
to indicate the start of the declining phase.

\section{Conclusions and Perspectives}\label{sec:Conclusions}

The basic physics of the solar tachocline began to be revealed to solar physicists with the seminal
paper of \citet{1992A&A...265..106S}. Since then, the physics of this
intriguing layer has led to many debates in the research community. In
particular, its extreme thinness continues to be a source of
questioning. The tachocline is a turbulent, doubly-sheared, highly stratified,
low-\Pra, and magnetized thin layer. It is a promising subject of
research for any fluid-dynamics enthusiast. We have seen in the
preceding sections that advanced, global 3D models could be developed
to address in a major way the dynamics of the tachocline. On the
other hands, such models are not resolved enough to model adequately
all the small-scale turbulent transport in the solar tachocline. In
that respect, several routes can be highlighted to continue deciphering the mysteries of the tachocline:

\begin{itemize}
\item In the past two years, the study of turbulent transport in the
  solar tachocline has been again revisited. On one hand,
  \citet{2020ApJ...901..146G} has developed in a clear manner the argument that the low-{\Pra}
  regime of the solar tachocline makes the estimate of the net
  turbulent transport non-trivial. On the other hand, the effect of
  magnetic field on the transport is still not fully
  understood. Following the important work of \citet{2007ApJ...667L.113T},
  \citet{2020ApJ...892...24C} have developed a theoretical argument for a
  complex tachocline confinement leveraging dynamo-generated tangled
  magnetic fields. We are today in the position of simulating the
  highly turbulent solar tachocline in a \FinalModif{more and more} realistic regime
  thanks to the massive development of 3D global numerical codes. 
  Dedicated, high-resolution numerical experiment could be built to address frontally this problem, e.g. taking into account the convective
  penetration in the upper tachocline and the associated magnetic
  field sustained by dynamo action above. 
\item The role of magnetoshear instabilities in the dynamics of the
  tachocline and of the solar dynamo is progressing actively recently. These
  have been found in some global models of the solar interior
  \citep{2015ApJ...813...95L,2019ApJ...880....6G,2020A&A...641A..13J}, and were interpreted to have
  different effects on the global dynamo. While the solar dynamo is responsible for generating the spot-producing magnetic fields and their migration in latitude, the global MHD of the tachocline provides a mechanism to create the longitudinal distribution of active regions. As discussed in section \ref{sec:NLE_3D}, there exists a major opportunity to link via data assimilation the observed spatio-temporal distribution of active regions with the simulated "imprints" of their latitude-longitude locations created by the global tachocline instabilities of the spot-producing magnetic fields (see, e.g. figure 12 of \citealt{2020SpWea..1802109D}).
\item The tachocline coincides at least in part with the
  convective-radiative transition in Sun-like stars. Angular
  momentum can be transported from one region to the other, especially
  when the star changes its structure or when the external convection
  zone spins down due to magnetic torques from its wind. In the theory of cool-stars'
  rotation, the coupling between these two regions plays an important
  role in the pre-main sequence and early main sequence
  phases. For instance, \citet{2013A&A...556A..36G} parameterized such angular
  momentum exchange timescale to reproduce the rotational
  distribution observed in open clusters (see also
  \citealt{2019A&A...621A.124B,2020A&A...635A.170A}). An assessment of this coupling timescale in realistic models of the tachocline \citep[see \textit{e.g.}][]{2011ApJ...742...79B} would therefore be extremely useful today to improve our understanding of the secular evolution of the Sun and solar-type stars.
\end{itemize}

\backmatter

\bmhead{Acknowledgments}

AS and ASB acknowledge partial financial support by DIM-ACAV+ ANAIS2 project, ERC Whole Sun Synergy grant \#810218 and STARS2 starting grant \#207430, ANR Toupies, ANR STORMGENESIS \#ANR-22-CE31-0013-01, INSU/PNST, Solar Orbiter and PLATO CNES funds, and GENCI via project 1623. MD acknowledges the support for this work from the National Center for Atmospheric Research, which is a major facility sponsored by the National Science Foundation under
cooperative agreement 1852977, and also acknowledges support from several NASA grants, 
namely NASA-LWS award number 80NSSC20K0355, NASA-HSR award number 80NSSC21K1676, subaward from JHU/APL's 
NASA-HSR grant with award number 80NSSC21K1678 and subaward from Stanford's COFFIES Phase II NASA-DRIVE Center with award number 80NSSC22M0162. Bernadett Belucz's work is supported by Newton International Fellowship of The Royal Society, program number NIF-R1-192417. GG acknowledges support from NASA grants
NNX14AB70G, 80NSSC20K0602, and 80NSSC20K1320.

\bmhead{Competing Interests} The authors declare they have no conflicts of interest.




\end{document}